\documentclass[12pt]{article}
\usepackage{a4}
\usepackage{amsfonts}
\usepackage{amssymb}
\usepackage{cite}
\usepackage{epsfig}
\usepackage[all]{xy}
\usepackage{amsmath}
\usepackage{cite}
\usepackage{lscape}

\newcommand{\Z}{\mathbb{Z}}

\newcommand{\bea}{\begin{eqnarray}}
\newcommand{\eea}{\end{eqnarray}}

\begin{document}

\title{
\begin{flushright} \vspace{-2cm}
{\small UPR-1174-T
}
\end{flushright}
\vspace{2.8cm}
Computation of D-brane instanton induced superpotential couplings -\\
Majorana masses from string theory \\
}
\vspace{2.5cm}
\author{\small Mirjam Cveti{\v c}, Robert Richter  and  Timo Weigand}

\date{}

\maketitle

\begin{center}
\emph{Department of Physics and Astronomy, University of Pennsylvania, \\
     Philadelphia, PA 19104-6396, USA } \\

\vspace{0.2cm}

\tt{cvetic@cvetic.hep.upenn.edu, rrichter@sas.upenn.edu, timo@sas.upenn.edu}
\vspace{1.1cm}
\end{center}
\vspace{1.0cm}

\begin{abstract}
\noindent We perform a detailed conformal field theory analysis of
$D2$-brane instanton effects in four-dimensional type IIA string
vacua with intersecting $D6$-branes. In particular, we explicitly
compute instanton induced fermion two-point couplings which play the
role of perturbatively forbidden Majorana mass terms for
right-handed neutrinos or MSSM $\mu$-terms. These results  can
readily be extended to higher-dimensional operators. In concrete
realizations of such non-perturbative effects, the Euclidean
$D2$-brane has to wrap a rigid, supersymmetric cycle with strong
constraints on the zero mode structure. Their implications for Type
IIA compactifications on the $T^6/({\mathbb Z}_2 \times {\mathbb
Z}_2)$ orientifold with discrete torsion are analyzed. We also
construct a local supersymmetric GUT-like model allowing for a class
of Euclidean $D2$-branes whose fermionic zero modes meet all the
constraints for generating Majorana masses in the phenomenologically
allowed regime. Together with perturbatively realized Dirac masses,
these non-perturbative couplings give rise to the see-saw mechanism.

\end{abstract}

\thispagestyle{empty}
\clearpage

\section{Introduction}

D-brane instantons in four-dimensional ${\cal N}=1$ supersymmetric
string compactifications have the potential to generate
perturbatively absent matter couplings of considerable
phenomenological importance
\cite{Blumenhagen:2006xt,Haack:2006cy,Ibanez:2006da, Florea:2006si}.
This mechanism relies on the existence of perturbative global
abelian symmetries of the  effective action which are the remnants
of $U(1)$ gauge symmetries broken by a St\"uckelberg type coupling.
Under suitable  conditions, non-perturbative effects can break these
global symmetries by inducing perturbatively forbidden since $U(1)$
charge violating effective couplings. Due to its phenomenological
relevance, the main focus of \cite{Blumenhagen:2006xt,Ibanez:2006da}
has been on the generation of a large Majorana mass term for
right-handed neutrinos. It has been shown that in principle, the
peculiar intermediate scale in the range $(10^8 - 10^{15})$ GeV of
this term can arise quite naturally and without drastic fine-tuning
from the genuinely stringy non-perturbative physics. Finding a
concrete realization of this scenario would therefore constitute an
interesting example of how string theory  provides a natural
explanation for an otherwise poorly understood origin of mass scales
in particle physics. Other important coupling terms potentially
generated in a similar manner include the hierarchically small
$\mu$-term of the MSSM
\cite{Blumenhagen:2006xt,Ibanez:2006da,Buican:2006sn} or the
Affleck-Dine-Seiberg superpotential of SQCD
\cite{Florea:2006si,Akerblom:2006hx}. Similar D-brane instanton
effects may also give rise to a realistic pattern of Yukawa
couplings \cite{Abel:2006yk} and play an important role in the
context of flux compactifications and de Sitter uplifting
\cite{Haack:2006cy}.

While both the presence of the mentioned perturbative abelian
symmetries and their non-perturbative breakdown are typical of many
different types of string constructions, they are particularly
obvious in the context of four-dimensional D-brane
models\footnote{Similar effects also arise in the S-dual heterotic
picture for compactifications with $U(n)$ bundles
\cite{Blumenhagen:2005ga,Blumenhagen:2005pm}.}. For definiteness let
us focus on Type IIA orientifolds involving intersecting D-branes
(see e.g.
\cite{Blumenhagen:2005mu,Blumenhagen:2006ci,Marchesano:2007de} for
up-to-date background material and references). Dual constructions
on the mirror symmetric Type IIB and Type I side have been studied
in \cite{Florea:2006si} (see also \cite{Buican:2006sn}) and
\cite{Bianchi:2007fx}, respectively. In our case, the relevant
non-perturbative objects are $E2$-instantons, i.e. Euclidean
$D2$-branes wrapping supersymmetric three-cycles of the internal
compactification manifold. These can induce superpotential couplings
of the form \bea \label{W1} W_{np}= \prod_i \Phi_i e^{-\frac{2
\pi}{{\ell}_s^3\, g_s}{\rm Vol}_{E2}} \eea between the chiral
charged matter superfields.

At the intersection of the $E2$-instanton wrapping the sLag $\Xi$
and a stack of ${N}_a$ $D6$-branes on, say, cycle $\Pi_a$, there exist
$[\Xi\cap \Pi_a]^+$ and $[\Xi\cap \Pi_a]^-$ fermionic zero modes
$\lambda_a$ and  $\overline \lambda_a$ in the fundamental and
antifundamental representation of $U(N_a)$, respectively. It follows
that the charge of the $E2$-instanton under the corresponding global
$U(1)_a$ symmetry is determined by the topological intersection of
the instanton cycle $\Xi$ with $\Pi_a$ and its orientifold image
$\Pi'_a$ as \cite{Blumenhagen:2006xt} \bea Q_a = {N}_a \, {\Xi}
\circ (\Pi_a - \Pi'_a). \eea As a result, the exponential
suppression factor in (\ref{W1}) can carry just the right global
$U(1)$ charge to cancel the charge of the operator $\prod_i \Phi_i$,
thus rendering the complete term  (\ref{W1}) invariant.

A detailed prescription for the conformal field theory computation
of the so induced interaction terms has been given in
\cite{Blumenhagen:2006xt}\footnote{For previous studies of the CFT
associated with the $D3-D(-1)$ system
 see
\cite{Green:2000ke,Billo:2002hm}. Note that $E2$-instantons at
general angles differ from this construction in that they do not
correspond to gauge instantons.}. Important constraints on the
instanton cycle $\Xi$ beyond the necessary condition of $U(1)$
charge cancelation arise from the requirement of absence of
additional uncharged instanton zero modes apart from the usual four
bosonic modes $x_E^{\mu}$  and their fermionic partners
$\theta^{\alpha}$, $\alpha =1,2$. In particular, the cycle $\Xi$ has to be
rigid so that there exist no reparametrization zero modes and it has
to satisfy $\Xi \cap \Xi' =0$ to avoid zero modes at the
intersection of $\Xi$ with its orientifold image. The presence of
additional zero modes is expected to give rise to higher fermion
couplings as opposed to contributions to the superpotential of the
form (\ref{W1}) (see \cite{Beasley:2005iu} for a discussion of such
terms arising from non-isolated worldsheet instantons in heterotic
$(0,2)$ models). The zero mode constraints are the main obstacle to the
construction of concrete string vacua featuring the described
non-perturbative couplings, and in fact no example of an
$E2$-instanton satisfying all of them has been given in the
literature so far. For an illustration of the associated challenges
in semi-realistic toroidal model building see \cite{Ibanez:2006da}. Besides
determining if these constraints can be realized at all, it is also
important to study if the scale of the induced couplings can indeed
account for the peculiar hierarchies associated to them in the MSSM.

The aim of this article is to
provide evidence that
these questions can be answered in the
affirmative.
We do so by constructing a local set-up of a toroidal intersecting
brane world  and  explicitly compute the $E2$-instanton induced Majorana mass terms.

For this purpose, we first need to continue and extend the study of
the CFT computation of $E2$-instanton effects initiated in
\cite{Blumenhagen:2006xt}. This includes, among other things, a
careful construction of the vertex operators for the twisted  zero
modes between the instanton and the $D6$-branes in section
\ref{sec_SUSY}. We then determine in section \ref{sec_Ampl-CFT} the
basic tree-level four-point amplitudes which are the building blocks
for non-perturbative couplings of the type (\ref{W1}). Special care
requires the analysis of the family replication structure. We
exemplify this point in section \ref{sec_Fam}.

In section 3, we apply our results to a concrete, but local model
meeting all criteria for the generation of $E2$-induced Majorana
mass terms.  While, unlike on general Calabi-Yau manifolds, the
underlying CFT of toroidal models is exactly solvable and a large
class of supersymmetric three-cycles is known explicitly, it is much
harder to construct rigid cycles in this framework as required for
the instanton sector.  The only known examples of rigid special
Lagrangians on toroidal backgrounds have been given in
\cite{Blumenhagen:2005tn} for the $\Z_2\times \Z_2$ orientifold with
torsion (called $\Z_2\times \Z_2'$ in the sequel) and, more
generally, in \cite{Blumenhagen:2006ab} for shift orientifolds. For the dual constructions with magnetized branes see \cite{Pradisi:2002vu,Dudas:2005jx}.

We construct a local supersymmetric $SU(5)$ GUT-like toy model on
the $\Z_2\times \Z_2'$ orientifold with right-handed neutrinos.
While its spectrum is far from realistic and fails to satisfy all global consistency conditions, the purpose of this setup
is to demonstrate that appropriate instanton sectors contributing to
the superpotential of the form (\ref{W1}) do exist in the geometric
regime as proposed in \cite{Blumenhagen:2006xt,Ibanez:2006da}. In fact, this is the first example even of a local model with these properties.
We classify the set of supersymmetric $E2$-instantons on factorisable cycles whose zero mode structure allows for superpotential two-fermion couplings of the above type.
There are  $64$ such rigid cycles lying on top of one of the orientifold planes and differing by their twisted charges.
Summing up their contributions gives rise
to Majorana masses for the right-handed neutrinos of the order of $10^{11}$ GeV.

\section{$D2$-brane instantons in Type IIA orientifolds}

In this section we summarize and provide additional details on the
computation of instanton corrections from Euclidean $D2$-branes
($E2$-branes). While the general framework has been described in
\cite{Blumenhagen:2006xt} (see also \cite{Ibanez:2006da}), subsection \ref{sec_SUSY} clarifies the
derivation of the instanton zero modes in some detail and contains a
careful construction of the associated vertex operators in toroidal
orientifolds. Subsection \ref{sec_Ampl-CFT} describes the exact CFT
computation of a prototype of instanton induced couplings, followed
by an illustrative example in subsection \ref{sec_Fam}.

\subsection{(Anti-)Instantons, zero modes and vertex operators}
\label{sec_SUSY}

Consider a Type IIA orientifold with  spacetime filling $D6$-branes
in the presence of a single Euclidean $D2$-brane wrapping the
three-cycle $\Xi$ of the internal manifold. In general, if  $\Xi$ is
not invariant under the orientifold action we have to consider also
the image $D2$-brane wrapping $\Xi'$.

For this topological sector to correspond to a local minimum of the
full string action, $\Xi$ has to be volume minimizing in its
homology class, i.e. special Lagrangian.

Recall that 
the concrete ${\cal N}=1$ subalgebra preserved by a
$D6$-brane wrapping the sLag $\Pi$ is determined by the
phase $\theta$ appearing in \bea \label{cal} {\rm{Im}}(e^{i \pi
\theta} \Omega|_{\Pi}) =0. \eea For $D6$-branes the value 
$\theta=0 \,\, {\rm mod}$ 2  corresponds to
the ${\cal N}=1$ algebra preserved also by the orientifold planes.
Standard arguments taking into account the localization of the
$E2$-brane in the external dimensions show that if it wraps a cycle
$\Xi$ with $\theta_{\Xi}=0$ it preserves the supercharges
$Q_{\alpha}, \overline Q'_{\dot \alpha}$ and breaks  $\overline Q_{\dot \alpha}, Q'_{\alpha}$, where the unprimed and primed quantities  
$Q_{\alpha},\overline Q_{\dot \alpha}$ and $Q'_{\alpha},\overline Q'_{\dot \alpha}$ generate the ${\cal N}=1$ subalgebra  preserved by the orientifold and the one orthogonal to it, respectively.

To restore
supersymmetry in the topological sector containing the $E2$-brane we
have to integrate all amplitudes over the corresponding Goldstone fermions
$\overline\theta_{\dot\alpha}$ and $\theta'_{\alpha}$ associated with the violation of
$\overline Q_{\dot \alpha}$ and  $Q'_{\alpha}$. Depending on the details of the orientifold projection, the $\theta'_{\alpha}$ can be projected out provided the cycle $\Xi$ is invariant, $\Xi=\Xi'$ \cite{LA}. In this case, the associated topological sector contributes to
the anti-holomorphic superpotential involving the anti-chiral
superfields. We identify it as the anti-instanton
sector. By contrast, the instanton, given by $\theta_{\Xi}=1 \,\,
{\rm mod}$ 2, preserves the  supercharges $\overline
Q_{\dot \alpha},  Q'_{\alpha}$ and violates $Q_{\alpha}, \overline
Q'_{\dot \alpha}$, thus contributing to the
holomorphic superpotential provided the $\overline\theta'_{\dot\alpha}$ are projected out. This is the situation we are interested in when computing
corrections to the superpotential.

In the sequel it will be useful to consider only 'aligned' cycles
$\Xi$ with $\theta_{\Xi}=0 \,\, {\rm mod}$ 2 wrapped by the
(anti-)instanton. At the CFT level, the fact that the instanton is
actually 'anti-aligned' w.r.t. the $D6$-branes internally is taken
into account by projecting the spectrum between the $E2$ and the
$D6$-branes of the model onto its GSO-odd part. For the
anti-instanton, we keep the GSO-even states. From the worldvolume
perspective, the two objects clearly carry opposite charge under the
RR three-form $C_3$ coupling to the worldvolume. The classical part
of the Euclidean (anti-)instanton action appearing as $e^{-S_{E2}}$
in corresponding F-term couplings reads \bea S_{E2} =  \frac{2
\pi}{\ell_s^3} \left( \frac{1}{g_s}{\rm Vol}_{\Xi} \mp i \int_{\Xi}
C_3 \right), \eea with the (lower) upper sign corresponding to
(anti-)instantons\footnote{We define the string length as
$\ell_s=\sqrt{2 \pi \alpha'}$.}.\\

The zero modes of the (anti-)instanton can be computed in setups
where the ${\cal N}=(2,2)$ CFT describing the internal sector is
known exactly. The general form of the various vertex operators can
be found in  \cite{Blumenhagen:2006xt}. For the sake of concreteness
and as preparation for our explicit computations, we specialize here
to the case that the $D6$- and the $E2$-branes wrap factorizable
three-cycles of toroidal orientifolds\footnote{A detailed summary of
the covariant open string quantization between two $D6$-branes in
this context can be found, e.g., in appendix A of
\cite{Cvetic:2006iz}.}.

There are three different sectors to distinguish corresponding to
the boundary conditions of the open strings:\\

$\bullet$ $E2-E2$: \\
This sector contains  the usual four bosonic zero modes $x_E^{\mu}$
corresponding to the Goldstone bosons associated with the breakdown
of four-dimensional Poincar\'e invariance due to the localization of
the instanton in spacetime. Their vertex operators (in the (-1)
picture) are simply given by \bea
 V_{x_E^{\mu}}(z)= \Omega_{E2E2}\, x_E^{\mu} \, \frac{1}{\sqrt{2}}\psi_{\mu}(z)\, e^{-\varphi(z)}.
\eea Here $\Omega_{E2E2}$ denotes the Chan Paton factor. The
polarization $x_E^{\mu}$ carries no mass dimension, corresponding to
a field in $d=0$ dimensions. It is related to the position
$x_0^{\mu}$ of the instanton in external spacetime via
$x_E^{\mu}=x_0^{\mu}/ \ell_s $. The factor $\frac{1}{\sqrt{2}}$
accounts for the fact that $\psi_{\mu}(z)$ are real fields. In this
and all following vertex operators we absorb the open string
coupling into the polarization (see section \ref{Ampl_gen} for a
detailed discussion of this point).

In general, the fermionic superpartners are given by four Weyl spinors
$\theta_{\alpha}, \overline \theta_{\dot\alpha}$  with vertex
operators (in the (-1/2) picture)
\bea
\label{vertex_theta}
 V_{\theta}(z) &=& \Omega_{E2E2}\, \theta_{\alpha} \,
     \, S^{\alpha}(z) \, \prod_{I=1}^3  e^{\frac{i}{2} H_I(z)} e^{-\varphi(z)/2}, \quad\quad\quad q =3/2,\nonumber \\
V_{\overline \theta}(z) &=& \Omega_{E2E2}\,
\overline\theta_{\dot\alpha} \,
     \, S^{\dot\alpha}(z) \, \prod_{I=1}^3  e^{-\frac{i}{2}\, H_I(z)} e^{-\varphi(z)/2},\quad\quad\,\, q =-3/2.
\eea Here $H_I(z)$ denotes the bosonization of the complexified
internal fermions $\Psi^I$ and $ S^{\alpha}$ ($ S^{\dot\alpha}$) are
the left-(right-)handed four-dimensional spin fields. We have also
included the worldsheet charge $q$. 
The fermionic zero modes are to be identified with the four Goldstinos discussed above (in (\ref{vertex_theta}) and in the sequel we omit the primes for simplicity).

Clearly, all these states are even
under the usual GSO-projection given in the covariant formulation by
\begin{align} \label{GSO}
 R: (-1)^F = (-i) \,\,  exp(i \pi \sum_{i=0}^4 s_i), \quad\quad
 NS:(-1)^F = (-1) \,\,  exp(i \pi \sum_{i=0}^4 s_i),
\end{align}
with $s_i = \pm 1/2$ and $\pm1$, respectively. As anticipated, in
computing amplitudes we have to integrate over the Goldstone bosons
$d^4 x_E^{\mu}$ as well as all four Goldstinos
$d^2\theta_{\alpha}$ and $d^2\overline \theta_{\dot\alpha}$ to restore
four-dimensional Poincar\'e invariance and ${\cal N}=1$
supersymmetry.
Only if the modes $\overline \theta_{\dot\alpha}$ ($\theta_{\alpha}$) are projected out \cite{LA} for (anti-)instantons  will this result in superpotential couplings.

In general the  $E2-E2$ sector also comprises $b_1(\Xi)$ chiral
superfields corresponding to the position moduli of the $E2$-brane. On
toroidal backgrounds, they are associated with the moduli along those
two-tori in which the $E2$-brane is not fixed. For completeness, we
display the vertex operators for the chiral component fields
corresponding to the position moduli in the, say, first torus, 
\bea
V_c(z) &=& \Omega_{E2E2}\, c\, e^{i H_1(z)} e^{-\varphi(z)}, \\
V_{\chi_{\alpha}} &=& \Omega_{E2E2}\, \chi_{\alpha}  S^{\alpha}(z)\,
e^{\frac{i}{2} H_1(z)} \prod_{I=2,3}  e^{-\frac{i}{2} H_I(z)} e^{-\varphi(z)/2}, 
\eea
to be supplemented in general by their anti-chiral counterparts, again possibly modulo the issue of orientifold projections. The need to
integrate in particular over the modulini yields non-vanishing
instanton amplitudes only once they are absorbed by couplings to
additional closed or open string fermionic modes. On toroidal
orbifolds, this results in higher fermion interactions\footnote{This
is in agreement with the analysis in \cite{Beasley:2003fx} of the
superpotential induced by non-rigid supersymmetric membrane
instantons on $G_2$ manifolds. The superpotential was found to be
proportional to the Euler characteristic of the moduli space of the calibrated
three-cycle wrapped by the instanton, which vanishes for non-rigid
factorizable sLags on toroidal backgrounds.} first discussed in the
context of worldsheet instantons for heterotic (0,2) models in
\cite{Beasley:2005iu}. The corresponding effect for non-rigid
heterotic five-branes has been studied in \cite{Buchbinder:2006xh}.
Being interested in contributions to the superpotential, we restrict
ourselves to the study of instantons wrapping appropriate rigid three-cycles
$\Xi$ in the sequel.
\\

$\bullet$ $E2-D6$ :\\
Additional zero modes arise at the intersection of the $E2$-brane with
the various $D6$-branes from open strings localized at the
intersection point (or the overlap manifold). Open strings in this
sector are subject to Dirichlet-Neumann boundary conditions in the
extended four dimensions and to mixed DN boundary conditions
internally, depending on the concrete intersection angles. The
external DN conditions shift the oscillator moding in these
directions by $1/2$. In the Ramond sector, the zero point energy is
still vanishing and we find massless fermions. The novelty as
compared to the case of spacetime filling branes at angles is that
the degeneracy of states is lifted in that the four-dimensional spin
fields $S_{\alpha}$ or $S_{\dot\alpha}$  are no longer present. This
also affects the details of the GSO-projection. In the NS sector,
the  vacuum energy is zero only for completely parallel branes,
which is the only situation with bosonic zero modes in the
non-singular geometric phase.

We first consider the case of non-trivial intersection of an
instanton (anti-instanton) wrapping $\Xi$ and a stack of $N_a$
$D6$-branes wrapping $\Pi_a$ in all three two-tori. It gives rise to
$[\Xi \cap \Pi]^+$ fermionic zero modes in the   $({\bf N_a},-1_E)$
($(1_E, {\bf \overline N_a})$) and $[\Xi \cap \Pi]^-$  fermionic
zero modes in the respective conjugate representation
\cite{Blumenhagen:2006xt}.

To see this the concrete form of the vertex operators needed. For
actual computations it is indispensable to carefully distinguish
between positive and negative intersection angles in the three
two-tori. Generically, the intersection number between factorizable
three-cycles $\Pi_a$ and $\Pi_b$ are given by \bea
I_{ab}=\prod^3_{I=1} I^{I}_{ab}\,\,, \eea where $I^I_{ab}$ denotes
the intersection number in the $I$-th torus. Here positive
(negative) intersection number $I^I_{ab}$ corresponds to positive
(negative) angle $\theta^I_{ab}$ and it is understood that
$|\theta^I_{ab}|<1$.

Given the total intersection number, say $I_{ab}>0$, one
distinguishes four different cases, three cases where one has
negative intersection number $I^I_{ab}$ in two internal tori and
positive in the left one and the symmetric one in which the
intersection numbers in all three internal two-tori are positive. In
supersymmetric configurations the intersection angles add up to 2
for the latter choice,  while for the other three their sum is $0$.

Consider now an instanton wrapping the cycle $\Xi$ such that all
intersection angles $\theta^I_{E2a}$ are positive for some cycle
$\Pi_a$ wrapped by a $D6$-brane. Upon projection onto states odd under
the GSO-operator (\ref{GSO}), the vertex operators for the fermionic
zero mode $\lambda_{a}$  at the intersection $E2-a$ is given by \bea
\label{vertex_la} V_{\lambda_{a}} =
 \Omega_{aE2} \,\lambda_a
\, \Sigma(z)\, \prod_{I=1}^3 \sigma_{1-\theta^I_{E2a}}(z)\,
e^{-i(\theta^{I}_{E2a}-\frac{1}{2}) H_I(z)} \,
e^{-\varphi(z)/2}.\eea

\noindent Here $\Sigma(z)$ denotes the bosonic twist field ensuring
Dirichlet-Neumann boundary conditions in spacetime. The $\lambda_a$
are Grassmannian variables and represent the polarization of the
fermionic zero mode, normalized again as a field in $D=0$
dimensions. Note that the GSO-projection forces us to keep only the
state in the sector starting from the $D6$-brane and ending on the
$E2$ and projects out the state with the reversed orientation. The
relevant intersection angles are therefore negative and lead to the
above form of the vertex (see e.g. \cite{Cvetic:2006iz}) carrying worldsheet
charge $q=-1/2$. As indicated by the CP indices, it transforms as
$({\bf N_a},-1_E)$.

For anti-instantons, we have to keep the state oriented from $E2$ to
$D6$ (i.e. transforming  as $(1_E, {\bf \overline N_a})$)  and of worldsheet
charge $q=1/2$.  We will refer to states negatively charged under
$U(1)_a$ as $\overline \lambda_a$. The various remaining cases are
dealt with analogously. This finally leads to the index theorem
stated above.

For completeness, we briefly discuss non-chiral intersections
between the E2- and the $D6$-branes. Consider first the
supersymmetric situation that the corresponding cycles are parallel
in one torus such that, say, $\theta^1_{E2a} >0$, $\theta^2_{E2a} =
-\theta^1_{E2a}$, $\theta^3_{E2a}=0$. For instantons, we find one
chiral fermionic zero mode in the $E2 \rightarrow D6_a$ sector with vertex \bea
V_{\overline\lambda_a}= \Omega_{ E2a} \,\overline\lambda_a \,
\Sigma(z)\,   \sigma_{\theta^1_{E2a}}(z)  e^{i(\theta^1_{E2a}-
\frac{1}{2} ) H_1(z)} \sigma_{1-\theta^1_{E2a}}(z)
e^{i(-\theta^1_{E2a}+\frac{1}{2}) H_2(z)} \, e^{-\frac{i}{2} H_3(z)}
\, e^{-\frac{\varphi(z)}{2}} \nonumber \eea and one in the $D6_a
\rightarrow E2$ sector with vertex \bea V_{\lambda_a}= \Omega_{aE2}
\,\lambda_a \, \Sigma(z)\,
 \sigma_{1-\theta^1_{E2a}}(z)  e^{i(-\theta^1_{E2a}+\frac{1}{2}) H_1(z)}
 \sigma_{\theta^1_{E2a}}(z)
 e^{i(\theta^1_{E2a}-\frac{1}{2}) H_2(z)} \, e^{-\frac{i}{2} H_3(z)} \,
 e^{-\frac{\varphi(z)}{2}}.
\nonumber  \eea Note that both zero  modes carry worldsheet charge
$q=-1/2$, i.e. are 'chiral' from the worldsheet point of view. The
corresponding modes for anti-instantons should be clear.

If finally the $E2$- and $D6$-brane are completely parallel
internally, the fermionic instanton zero mode sector for non-rigid
cycles comprises simply the four states \bea V_{\lambda_a}=
\Omega_{ aE2} \,\lambda_a  \, \Sigma(z)\,\prod_{I=1}^3  e^{i s_I
H_I(z)} \, e^{-\varphi(z)/2} \eea with worldsheet charge $q=\sum_I s_I=3/2$
or $-1/2$ and likewise four $\overline\lambda_a$ in the $E2
\rightarrow D6_a$ sector. Note that for completely rigid branes only
the two states $q=3/2$ are present. Since the zero point energy
vanishes for completely trivial intersections, the lowest lying
bosons are now also massless. In both the $E2 \rightarrow D6_a$ and
the $E2 \rightarrow D6_a$  sector the GSO-projection removes 2 out
of the 4 spinorial groundstates from the external dimensions, leaving
for instantons \bea V_{w_{\dot\alpha}} = \Omega_{ aE2} \, w_{\dot\alpha}
S^{\dot\alpha}(z) \Sigma(z) \, e^{-\varphi(z)} \eea (plus the
orientation reversed one), whereas the anti-instanton carries
chiral modes.
\\

$\bullet$ $E2-E2'$: \\
In general, there are additional zero modes at the intersection of
the $E2$-brane and its orientifold image. Due to the
Dirichlet-Dirichlet boundary conditions, the orientifold projection
picks up an additional minus sign as compared to the $D6-D6'$
sector. For single instantons, we therefore find $\frac{1}{2} (
I_{E2 E2'} + I_{O6 E2})$ bosonic and/or fermionic zero modes carrying charge $2$ under
$U(1)_{E2}$. Their vertex operators are identical to the more familiar massless states between
$D6$-branes and their images.  
For a straightforward generation of actual
superpotential terms we insist that these modes be absent.

\subsection{Amplitudes - Generalities and normalisation}
\label{Ampl_gen}

$E2$-instantons of the above kind can induce F-term couplings
involving the open string superfields $\Phi_{ab}$ between the
$D6$-branes present in the model. Of particular interest are those
couplings which are absent perturbatively since they violate some of
the global abelian symmetries which are the remnants of the $U(1)$
gauge symmetries on the $D6$-branes broken by St\"uckelberg-type
couplings to the RR-forms of the background. The exponential
suppression factor $e^{-S_{E2}}$ characteristic for instantonic
couplings transforms under the global $U(1)$ symmetries in such a
way that the full coupling \bea W_{np} = \prod_i \Phi_{a_i b_i}\,
e^{-S_{E2}} \eea is invariant again. More precisely, from the
axionic shift symmetries under these abelian symmetries induced by
the Chern-Simons couplings of the $N_a$ $D6_a$-branes one finds the
$U(1)_a$ transformation of the instanton \cite{Blumenhagen:2006xt} (see also \cite{Ibanez:2006da}),
\bea
 e^{-S_{E2}}=\exp\left[ \frac{2\pi}{ \ell_s^3}
           \left( -\frac{1}{g_s} {\rm Vol}_{\Xi} + i \int_{\Xi} C^{(3)}
         \right) \right] \longrightarrow  e^{i\, Q_a(E2)\,\Lambda_a}  \,\, e^{-S_{E2}}
\eea
with
\bea
\label{chargee}
          Q_a(E2)={N}_a\,\, \Xi\circ (\Pi_a - \Pi'_a).
\eea Indeed this charge is exactly the amount of $U(1)_a$ charge
carried by the fermionic zero modes between the $E2$ and the $D6_a$,
which serves as an important check that our identification of the
instanton vs. anti-instanton and the associated choice of
GSO-projection is correct.

The general procedure for the computation of the instanton induced
physical $M$-point couplings involving the canonically normalized
fields of the four-dimensional effective action
 has been outlined in  \cite{Blumenhagen:2006xt}. In
momentum space it is given, after some refinements, by \bea
\label{ampl_gen1} && \langle \Phi_{a_1,b_1}(p_1)\cdot\ldots\cdot
\Phi_{a_M,b_M}(p_M)
\rangle_{E2-{\rm inst}} = \nonumber \\
&& 
-\frac{1}{C} \int d^4 \tilde x_E \, d^2 \tilde \theta \,\,
       \sum_{\rm conf.}\,\,  {\textstyle
  \prod_{a} \bigl(\prod_{i=1}^{ [\Xi\cap
             \Pi_a]^+}  d \tilde\lambda_a^i\bigr)\,
               \bigl( \prod_{i=1}^{ [\Xi\cap
             \Pi_a]^-}  d\widetilde {\overline\lambda}_a^i\bigr) } \ \,\, e^{-S_{E2}} \,
         \times \, e^{Z'_0} \,  \nonumber \\
&&\phantom{aaa}
\times \langle \widehat\Phi_{a_1,b_1}[\vec x_1]   \rangle_{\lambda_{a_1},\overline{\lambda}_{b_1}}\cdot
            \ldots \cdot  \langle \widehat\Phi_{a_L,b_L}[\vec x_L]
          \rangle_{\lambda_{a_L},\overline{\lambda}_{b_L}} \,\times \,
         \prod_k \langle \widehat\Phi_{c_k,c_k}[\vec x_k] \rangle^{\rm
           loop}_{A(E2,D6_{c_k})}  . \nonumber
\eea Its basic building blocks are disk and annulus diagrams with
insertion of an appropriate product of boundary changing vertex
operators, denoted schematically by $ \widehat\Phi_{a_1,b_1}[\vec
x_1]$. It is understood that either precisely two of these diagrams
carry one $\theta$-mode each or one of them carries both. Each disk
carries two of the fermionic modes $\lambda_a$ from the $E2-D6$
sector, whereas the annulus diagrams are uncharged in that they are
free of $\lambda_a$-insertions. The instanton suppression factor
$e^{-S_{inst}}$ arises from exponentiation of tree-level disks with
no matter insertion and is corrected by the exponentiated
regularized one-loop amplitude $e^{Z'}$ with \bea \label{1loop} Z' =
\sum_a \left[ {\cal A}(D6_a,E2)+{\cal A}(D6'_a,E2)\right] + {\cal
M}(E2,O6) \eea in terms of the annulus and M\"obius amplitudes
${\cal A}$ and ${\cal M}$ (for details see
\cite{Blumenhagen:2006xt}). This one-loop factor has been computed
in \cite{Abel:2006yk} and \cite{Akerblom:2006hx} and is related to the regularized threshold
correction to the gauge coupling of a $D6$-brane wrapping the same
internal cycle $\Xi$ as the instanton. The latter have been determined in \cite{Lust:2003ky} for toroidal orientifolds. The combinatorical prefactor $-\frac{1}{C}$
arises from expansion of the exponentiated instanton moduli action $e^{-S_{mod}}$ containing the tree-level and
annulus coupling terms.
\\

After this review we clarify the proper $g_s$-normalization of the
instanton amplitude. It is convenient to work in the frame where all
vertex operators (including the ones for fields between two
$D6$-branes) carry no explicit factors of $g_s$. A disk with boundary
only on one type of $Dp$-brane carries the normalization factor
 \bea C_p=
\frac{2\pi}{g_s \ell_s^{p+1}}. \eea Consequently, all kinetic and
tree-level perturbative coupling terms arise formally at order
$g_s^{-1}$. This is therefore the tree-level order in $g_s$ to which
non-perturbative couplings are to be compared.

The disks appearing in the above expression (\ref{ampl_gen1}) are
bounded partly by the $D6$-branes $a$ and $b$ and the
$E2$-instanton. In such a case, the amplitude has to be normalized
with respect to the dimension of the overlap of the branes involved
and therefore carries a factor (see also \cite{Akerblom:2006hx})
\bea \label{norm_disk} C = \frac{2 \pi}{g_s}. \eea

\noindent Consider  now the normalization of the instanton moduli
measure. As noted, the integration over the four-dimensional
supermoduli $ \int d^4 \tilde x_E \, d^2 \tilde \theta$ restores
Poincar\'e invariance and ${\cal N}=1$ supersymmetry. The inclusion
of the charged zero modes $\tilde \lambda_a$ in the measure can be
understood as the process of integrating these modes out since they
would result in a zero in the Pfaffian $e^{Z'}$
\cite{Witten:1999eg}. While the Grassmannian integral is trivial and
merely results in a combinatorical factor, the integration over $d^4
\tilde x_E$ will ensure momentum conservation of the $M$-point
amplitude (see equ. (\ref{dxint})). The tilde indicates that we have
to integrate over the properly normalized zero modes corresponding
to the instanton moduli in the ADHM action in the limit where the
$E2$-brane wraps the same cycle as one of the $D6$-branes and
therefore represents a gauge instanton (or its stringy
generalization for $\ell_s \neq 0$.) The resulting Jacobian in the
transition from the polarizations appearing in our vertex operators
takes care of the proper normalization procedure in the more
familiar case of field theory instantons (see e.g. the review
\cite{Shifman:1999mv} for details). The situation of parallel $E2$
and $D6$-branes is T-dual to the $D(3)-D(-1)$ system in Type IIB
theory. Adapting the CFT analysis of \cite{Billo:2002hm} to our
case\footnote{Unlike \cite{Billo:2002hm} we do not assign
four-dimensional canonical mass dimensions to the instanton moduli
but treat them as dimensionless fields in zero dimensions. The disk
normalisation between parallel $E2$ and $D6$-branes is $2 \pi
{\cal V}_{E2}/{g_s}$. The resulting amplitudes and effective moduli action
before rescaling therefore differ by a power of ${\cal V}_{E2}/\ell_s^4$ as
compared to the ones in \cite{Billo:2002hm}. Our rescaling for the
case of parallel $E2$ and $D6$ systems is otherwise identical upon
replacing $g_0 \to \sqrt{g_s \,{\cal V}_{E2}/\pi}$. Finally, for $E2$ and
$D6$-branes at angles, the rescaling of the $\lambda$-modes does not
contain any ${\cal V}_{E2}$, in agreement with (\ref{norm_disk}).}, we find
the following relation between the polarization in the vertices and
the ones to appear in the measure, \bea \label{scaling} \tilde
x_E^{\mu} = \frac{x^{\mu}_E}{2} \sqrt{\frac{2 \pi {\cal V}_{E2}}{g_s}},
\quad\quad \tilde \theta^{\alpha} = \theta^{\alpha}  \sqrt{\frac{2
\pi {\cal V}_{E2}}{g_s}} \quad\quad \tilde \lambda = \lambda  \sqrt{\frac{2
\pi}{g_s}}, \label{transformation}\eea where ${\cal V}_{E2} = {\rm {Vol}}_{E2}/
\ell_s^3$. Most importantly, the contribution from two
$\lambda$-modes\footnote{Recall that $d(a \psi) = a^{-1} d \psi$ for
a Grassmann field $\psi$.} cancels the $g_s$-dependent topological
normalization of the disk (\ref{norm_disk}).

If indeed precisely  2 $\lambda$-modes are inserted per disk (and
none on the annulus diagrams carrying an additional factor of $g_s$
in their normalization), then the induced M-point amplitude is
proportional to $\frac{2 \pi {\cal V}_{E2}}{g_s}$ due to the remaining
normalization factors from $ \int d^4 \tilde x_E \, d^2 \tilde
\theta$ (times the exponential dependence, of course). It therefore
arises at 'string tree-level' as compared to the perturbative terms.

\subsection{Amplitudes - CFT details}
{\label{sec_Ampl-CFT}

A phenomenologically interesting application is the computation of
instanton-induced $U(1)$ charge violating 2-point couplings. These
can be thought of as Majorana masses for right-handed neutrinos or
as $\mu$-terms in the MSSM
\cite{Blumenhagen:2006xt,Ibanez:2006da,Buican:2006sn}. We will now
compute such 2-point couplings in a general setup, which can then be adapted to concrete examples.

Consider the superfield $\Phi^A_{ab}$ at the intersection $A$
between two $D6$-branes wrapping the cycles $\Pi_a$ and $\Pi_b$. We
would like to generate couplings of the form $<\psi^A_{ab}
\psi^B_{ab}>_{E2}$. The zero mode structure of the instanton has to
allow for a compensation of the excess of $U(1)_a$ and $U(1)_b$
charge. This requires

\begin{align}
\label{intnum_e2}
[\Pi_{E2} \cap \Pi_{a}]^+=2 &\quad&  [\Pi_{E2} \cap \Pi_{b}]^-=2 &\quad & {\rm for} &\quad & I_{ab} >0 \nonumber\\
[\Pi_{E2} \cap \Pi_{a}]^-=2 & \quad &[\Pi_{E2} \cap \Pi_{b}]^+=2 &\quad & {\rm for} &\quad & I_{ab} <0
\end{align}

\noindent and the intersection between the $E2$ and all other
$D6$-branes has to vanish\footnote{\label{importantfootonte}Strictly
speaking, this is only true if the $E2$ lies away from the
orientifold brane. In case $E2=E2'$, the $E2-a$ and $a'-E2$ sector
are identified.}. We reiterate that the absence of additional
reparametrization and other uncharged zero modes in the $E2-E2'$
sector necessitates the $E2$-brane to be rigid and to satisfy
$[\Pi_{E2} \cap \Pi_{E2'}]^{\pm} =0$. The four zero modes are
denoted by $\lambda_a^i$ and $\overline \lambda_b^k$ for $i,k =
1,2$. Since the CFT computation depends on the concrete form of the
vertex operators, we have to make a definite choice of angles and
intersection numbers. Consider e.g. the simple situation
corresponding  to $I_{ab} >0$ such that\footnote{Note that in most
concrete realizations including our example given in section
\ref{sec_model} the angles will be less symmetric, but this can
easily be dealt with.} \bea
&&\theta^I_{ab}>0, \quad \theta^I_{E2a}>0,  \quad  \theta^I_{E2b}<0, \nonumber\\
 &&\sum^3_{I=1}\,\theta^I_{ab} = \sum^3_{I=1}\, \theta^I_{E2a} =2 = -\sum^3_{I=1}\, \theta^I_{E2b}. \label{anglesum}
\eea
With this choice of angles the vertex
operator for $\psi^A_{ab}$ takes the form
 \bea
\label{vertex_psi} V_{\psi^A_{ab}} = \ell_s^{\frac{3}{2}}\,
\Omega_{ba}\, \psi^A_{\alpha} \, S^{\alpha}(z) \, \prod_{I=1}^3
\sigma_{1-\theta^I_{ab}}(z)\, e^{-i(\theta^I_{ab}-1/2) H_I(z)}\,
e^{i k^A_{\mu}X^{\mu}(z)} \,e^{-\varphi(z)/2}, \, \eea where
$\psi^A_{\alpha}$ carries canonical mass dimensions. The vertex for
the zero mode at the intersection of $E2$ and $D6_a$ has been given
in (\ref{vertex_la}), and the one between $E2$ and $D6_b$ reads \bea
\label{vertex_lb} V_{\overline\lambda^k_{b}} = \Omega_{E2b} \,
\overline\lambda^k_b
 \, \Sigma(z)\, \prod_{I=1}^3
\sigma_{1+\theta^I_{E2b}}(z)\, e^{i(\theta^{I}_{E2b}+1/2)
H_I(z)}\,e^{-\varphi(z)/2}\,\,.\eea We then have to compute

\bea \label{2point1} <\psi^A_{ab} \psi^B_{ab}>_{E2} &=& -\frac{1}{2!}\,\frac{{\cal V}_{E2}}{16}
\frac{g_s}{2\pi}\,\int d^4  x_E \,\int d^2  \theta \, \int d^2
\lambda_a\, \int d^2 \overline\lambda_{b}\, \, e^{-S_{inst.}}\,
e^{Z'} \nonumber
\\ && \sum_{i,j,k,l} < V^{-\frac{1}{2}}_{\Theta^{\alpha}}  V^{-\frac{1}{2}}_{\overline\lambda^k_{b}} V^{-\frac{1}{2}}_{\psi^A_{\alpha}} V^{-\frac{1}{2}}_{
\lambda^i_a}
>\,< V^{-\frac{1}{2}}_{\Theta^{\beta}}  V^{-\frac{1}{2}}_{\overline\lambda^l_{b}} V^{-\frac{1}{2}}_{\psi^B_{\beta}} V^{-\frac{1}{2}}_{
\lambda^j_a}
>.
\eea This already includes the rescaling \eqref{scaling}. It is
understood that the summation is only over those combinations of
family indices with non-trivial disk diagrams. This important point
has to be studied in concrete examples.

The disk amplitudes appearing in (\ref{2point1}) can be evaluated
using standard CFT methods. The  computation of the four-point
function $<\theta \, \overline\lambda \psi \, \lambda>$ involving
the vertex operators
(\ref{vertex_theta},\ref{vertex_la},\ref{vertex_psi},\ref{vertex_lb})
requires the following correlators \begin{align} \nonumber
<\prod^4_{i=1}\,e^{-\varphi(z_i)/2}>\,=\,\prod^{4}_{i=1}z^{-\frac{1}{4}}_{ij},
\qquad \qquad &<S^{\alpha}(z_1)\,
S^{\beta}(z_2)>\,=\,\epsilon^{\alpha\beta}z^{-\frac{1}{2}}_{12},
\qquad
\\
<e^{i\alpha H_I(z_1)}\, e^{i\beta H_I(z_2)}\, e^{i\gamma H_I(z_3)}
\, e^{i\delta H_I(z_4)}>\,&=\, z^{\alpha\,\beta}_{12}\,
z^{\alpha\,\gamma}_{13}\,
z^{\alpha\,\delta}_{14}\,z^{\beta\,\gamma}_{23}
\,z^{\beta\,\delta}_{24}\,z^{\gamma\,\delta}_{34},
\\
 \qquad \qquad \qquad <\Sigma(z_1)\,  e^{i k_{\mu}X^{\mu}(z_2)} \, \Sigma(z_3)>\,&=\,
e^{ik_{\mu}x^{\mu}_0}\, z^{-\frac{1}{2}}_{13}. \qquad \qquad \qquad
\qquad \qquad \qquad \nonumber
\end{align} Here $z_{ij}$ denotes $z_i-z_j$ and $x_0$
is the position of the E2-instanton in spacetime. The most involved
ingredient is the correlator of the three bosonic twist fields. In
general, it reads \cite{Cvetic:2003ch,Lust:2004cx} \bea <\sigma_{\alpha}(z_1)
\sigma_{\beta}(z_2) \sigma_{\gamma}(z_3)>\,= \left( 4\pi\,
\Gamma_{\alpha,\,\beta,\,\gamma} \right)^{\frac{1}{4}}
z^{-\alpha\,\beta}_{12} z^{-\alpha\,\gamma}_{13}
z^{-\beta\,\gamma}_{23} \sum_{m} e^{-{\cal A}(m)}, \label{bosonic
twist correlator} \eea where $\Gamma_{\alpha,\,\beta,\,\gamma}$ is
given by \bea \Gamma_{\alpha,\,\beta,\,\gamma}=
 \frac{\Gamma(1-\alpha)\,\Gamma(1-\beta)\,
\Gamma(1-\gamma)}{\Gamma(\alpha)\Gamma(\beta) \Gamma(\gamma)} \eea
and ${\cal A}(m)= A(m)/(2 \pi \alpha')$ is the area in string units
of the triangle formed by the three intersecting branes. The
correlator \eqref{bosonic twist correlator} vanishes if the angles
$\alpha$, $\beta$, $\gamma$ do not add up to an integer. The
normalization $(4\pi)^{\frac{1}{4}}$ was determined in
\cite{Cvetic:2003ch} by factorizing the four-point amplitude
involving four bosonic twist fields in the limit corresponding to a
gauge boson exchange.

Putting everything together, including the disk normalization factor
$\frac{2\pi}{g_s}$ and using the supersymmetry conditions
(\ref{anglesum}) result in
\bea
< \theta^{\alpha} \overline \lambda^k_b    \psi^A_{\alpha} \,\lambda^i_a\, \,
>  = \frac{2\pi \,
\ell^{\frac{3}{2}}_s}{g_s}Tr\left( \Omega_{E2E2}\Omega_{E2b}
\Omega_{ba}\Omega_{aE2} \right) \, \theta^{\alpha}
\,\overline\lambda^k_b \psi^{A}_{\alpha}\, \lambda^i_a \, \qquad
\qquad \qquad
\\ \times
\prod^3_{I=1}\left[
4\pi\Gamma_{1-\theta^I_{ab},\,1-\theta^I_{E2a},\,1+\theta^I_{E2b}}\right]^{\frac{1}{4}}
\sum_{m_I}e^{-{\cal A}^{A}_{ik}(m_I)} \int
\frac{\prod^4_{i=1}dz_{i}}{V_{CKG}}\,
z^{-\frac{1}{2}}_{13}\,z^{-\frac{1}{2}}_{24}\prod^{4}_{i,j=1} \,
z^{-\frac{1}{2}}_{ij}\, \,e^{-ik^A_{\mu}x^{\mu}_0}. \nonumber \eea
After we fix the vertex operator positions to\footnote{We need to
include the other cyclic order as well.} \bea z_1=0, \qquad z_2=x,
\qquad z_3=1, \qquad z_4=\infty \eea
 and add the
c-ghost part \bea <c(z_1)\,c(z_3)\, c(z_4)>=z_{13}\,z_{14}\,z_{34}
\eea the amplitude computes to\footnote{Note that even after taking
into account the other cyclic order the only non vanishing trace is
$Tr\left(\Omega_{E2E2}\Omega_{E2b} \Omega_{ba} \Omega_{aE2}\right)$
and therefore we need to integrate from $0$ to $1$.}
\begin{align}
< \theta^{\alpha} \overline \lambda^k_b    \psi^A_{\alpha} \,\lambda^i_a\, \,
>\,&= \frac{2\pi}{g_s}\,\ell_s^{\frac{3}{2}}\,
C^A_{ik}\, \,e^{-ik^A_{\mu}x^{\mu}_0} \quad (\theta^{\alpha} \overline \lambda^k_b    \psi^A_{\alpha}
\,\lambda^i_a )\, \label{buildingblock1}
\end{align}
with
\begin{align}
\label{C-const}
C^A_{ik}= \pi\, \prod^3_{I=1}  \left[ 4\pi\Gamma_{1-\theta^I_{ab},
\,1-\theta^I_{E2a},\,1+\theta^I_{E2b}}\right]^{\frac{1}{4}}
\sum_{m_I}e^{-{\cal A}^{A}_{ik}(m_I)}.
 \end{align}
Here we omit the trivial trace structure and use \bea \int^1_0
dx\left[x(1-x)\right]^{-\frac{1}{2}}=\pi. \eea In order to obtain the
coupling  $<\psi^{A}_{ab} \psi^{B}_{ab}
>_{E2}$ we plug \eqref{buildingblock1}
into \eqref{2point1} and perform the integrals over all fermionic
and bosonic zero modes. In doing so, we make use of the integral
representation of the $\delta$-function (recall that $x_0^{\mu} = \ell_s x_E^{\mu}$), \bea
\label{dxint}
\int d^4 x_E \,
e^{-ik_{\mu}x_0^{\mu}}\,= \frac{(2\pi)^4}{\ell^{4}_s}\,
\delta^{4}(\,k\,)\,\,, \eea and find
\bea
\label{bilinear_final}
<\psi^{A}_{ab} \psi^{B}_{ab}>_{E2}\, &=& -\frac{{\cal V}_{E2} \,\pi}{16\,g_s}\,M_s \,e^{-S_{inst.}} e^{Z'} \, \, \psi^{A}_{\alpha}  \epsilon^{\alpha \beta}\psi^{B}_{\beta} \quad\quad\quad\quad \quad\quad\,\,\quad\quad\quad\quad \quad\quad \nonumber\\
&& \qquad\qquad\qquad \times (2\pi)^4\,\delta^{4}\left(k^A+k^B\right)\,\sum_{i,j,k,l}
 \epsilon^{ij}\epsilon^{kl} \,C^A_{ik}C^B_{jl}.
\eea
The overall sign can always be absorbed into phases of the fermions.
Note that due to the Grassmannian integral, non-vanishing mass terms occur only for a  suitable family structure such that indeed $\sum_{i,j,k,l}
 \epsilon^{ij}\epsilon^{kl} \,C^A_{ik}C^B_{jl} \neq 0$.

\subsection{ Family structure for (off-)diagonal bilinears in a simple example}
\label{sec_Fam}

\begin{figure}
\begin{center}
\includegraphics[scale=0.50]{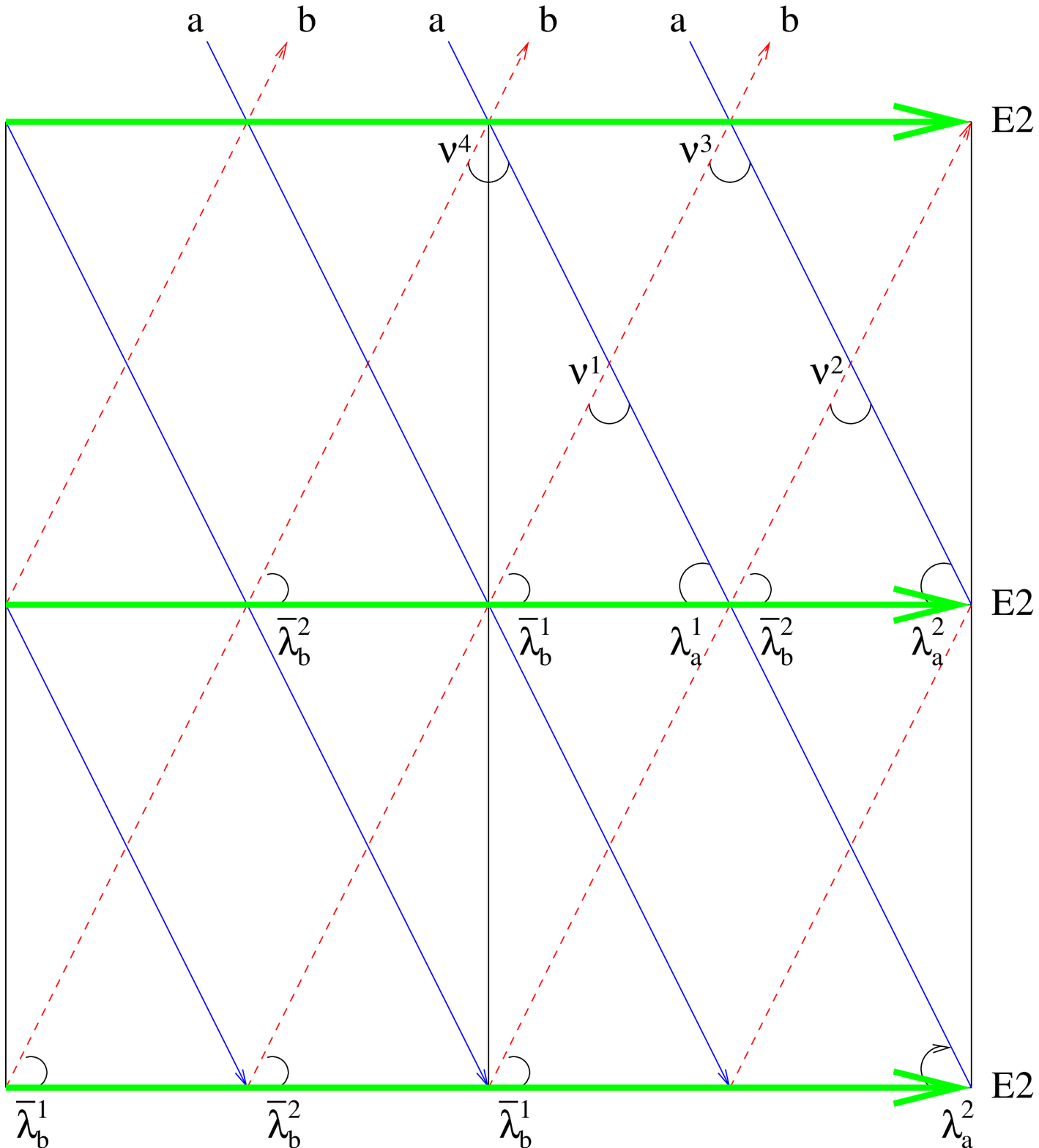}
\caption{Example of family structure yielding diagonal and off-diagonal fermion bilinears.} \label{fig_4neutrino}
\end{center}
\end{figure}

To appreciate this latter point, consider the family structure
depicted schematically in figure \ref{fig_4neutrino}. We can  think
of it as representing one of the three two-tori of a toroidal
orientifold model with factorizable $D6$-branes wrapping a one-cycle
on each torus. In this case, branes $a$ and $b$ correspond to
wrapping numbers $(1,-2)$ and $(1,2)$, respectively.  For simplicity
we assume  here that the complete family replication is due to
multiple intersections on just the depicted torus, though more
general situations leading to non-vanishing terms on a factorizable
$T^6$ are possible\footnote{In fact, our local model discussed in
the next section is more general in this respect.}. Consider now the
instanton wrapping the cycle $(1,0)$.

Straightforward inspection of the possible triangles connecting the
various intersection points reveals that the coupling terms in the
(zero-dimensional) moduli action of the instanton are proportional
to \bea \label{couplS_inst} &&\nu^1 \theta \, \overline\lambda_b^1
\, \lambda_a^1 \,\, e^{-{\cal A}^1_{11}} + \nu^1 \theta \,
\overline\lambda_b^2 \,\, \lambda_a^2 \, e^{-{\cal A}^1_{22}}
 + \nu^2 \theta \, \overline\lambda_b^2 \, \lambda_a^2 \,\, e^{-{\cal A}^2_{22}} + \nu^2 \theta \, \overline\lambda_b^1 \, \lambda_a^1 \,\, e^{-{\cal A}^2_{11}} + \nonumber \\
&& \nu^3 \theta \, \overline\lambda_b^2 \, \lambda_a^1 \,\, e^{-{\cal A}^3_{21}} + \nu^3 \theta \, \overline\lambda_b^1 \, \lambda_a^2 \,\, e^{-{\cal A}^3_{12}}
 + \nu^4 \theta \, \overline\lambda_b^1 \, \lambda_a^2 \,\, e^{-{\cal A}^4_{12}} + \nu^4 \theta \, \overline\lambda_b^2 \, \lambda_a^1 \,\, e^{-{\cal A}^4_{21}}.
\eea Note that we have only given the leading area suppression due
to the smallest possible triangles. The Grassmann integration now
dictates the possible combination of fermion bilinears appearing the
amplitude. This results in the following fermion bilinears in the
four-dimensional effective action, \bea S_{non-pert.} = - \int d^4x
\, \, \frac{2 \pi}{g_s} \, M_s \, e^{-S_{inst}} e^{Z'} \, \, \,
(\nu^1 \nu^2  \nu^3 \nu^4) \, \, {\cal M} \, \,  (\nu^1 \nu^2  \nu^3
\nu^4)^T \eea with the $4 \times 4$ matrix
\[ {\cal M} =  \left( \begin{array}{cc}
A & 0 \\
0 & B
\end{array} \right)\]
given in terms of
\[ A = \frac{\pi^2}{16} (4 \pi \Gamma)^{1/2} \left( \begin{array}{cc}
e^{-(\alpha + \beta)} & \frac{1}{2} ( e^{-2 \alpha} +  e^{-2 \beta}) \\
\frac{1}{2} ( e^{-2 \alpha} +  e^{-2 \beta})  & e^{-(\alpha + \beta)}
\end{array} \right),  \] \\
\[B = \frac{\pi^2}{16} (4 \pi \Gamma)^{1/2}
\left( \begin{array}{cc}
e^{-(\gamma + \delta)} & \frac{1}{2} ( e^{-2 \gamma} +  e^{-2 \delta}) \\
\frac{1}{2} ( e^{-2 \gamma} +  e^{-2 \delta})  & e^{-(\gamma + \delta)}
\end{array} \right). \]
Here we have defined
\bea
&&{\cal A}^1_{11} = {\cal A}^2_{22} = \alpha, \quad \quad {\cal A}^1_{22} = {\cal A}^2_{11} = \beta, \nonumber \\
&&{\cal A}^3_{21} = {\cal A}^4_{12} = \gamma, \quad \quad {\cal
A}^3_{12} = {\cal A}^4_{21} = \delta \label{areas} \eea and we have
omitted the $(2 \pi)^4 \delta (\sum k)$ in going from momentum to
position-space. We have also made use of our freedom to absorb a
phase $e^{i \pi/2}$ into the fields $\nu^1$ and $\nu^2$ to adjust
the signs of the Majorana mass terms.

As a result, we have found both diagonal couplings $\nu^i \nu^i$ and
the off-diagonal ones $\nu^1 \nu^2$ and $\nu^3 \nu^4$. The overall
scale of these terms is governed by the exponential suppression
factor $e^{-S_{inst}}$, whereas the relative size of the various
couplings is set by the ratio of the triangles involved. These
depend on the concrete K\"ahler and open string moduli. For example,
for a particular choice of brane positions we can set one of the
areas, say $\alpha$, to zero, in which case the off-diagonal
coupling $\nu^1 \nu^2$ would dominate over the diagonal ones $\nu^1
\nu^1$, $\nu^2 \nu^2$.

Finally we point out that the above non-perturbative couplings in
this example are allowed since not all possible intersection points
are connected by worldsheet instantons, i.e. disk triangles. As
observed already in \cite{Cremades:2003qj}, this is a generic
consequence of the fact that the three intersection numbers
$I_{Ea}$, $I_{Eb}$, $I_{ab}$ are not coprime. If, by contrast, in
addition to the couplings (\ref{couplS_inst}), also the combination,
say, \bea \nu^1 \theta \, \overline\lambda_a^1 \, \lambda_b^2 \,\,
e^{-\widetilde{\cal A}^1_{12}} + \nu^1 \theta \,
\overline\lambda_b^2 \lambda_a^1 e^{-\widetilde{\cal  A}^1_{21}}
\eea were present, the Grassmann integral would give zero for the
coupling $\nu^1 \nu^1$ whenever $\widetilde{\cal
A}^1_{11}+\widetilde{\cal A}^1_{22} = {\cal A}^1_{12}+{\cal
A}^1_{21}$.  This results in yet another important constraint on the
architecture of concrete models exhibiting $E2$-instanton effects, as has also been addressed in \cite{Ibanez:2006da}.

\section{Non-perturbative Majorana masses in a local GUT-like brane set-up}
\label{sec_model}

In this section we present a local brane configuration on the
orientifold $T^6/\Z_2\times \Z_2'$ which serves as a toy model for
realizing the see-saw mechanism for neutrino masses. While our
ultimate object of desire are globally consistent MSSM-like string
vacua satisfying all tadpole- and K-theory constraints, we content
ourselves for the time being with a local model with GUT gauge
group.  Apart from demonstrating the CFT techniques developed in the
previous section, our primary aim is two-fold: First to show that
rigid cycles meeting the strong requirements for the generation of
2-point couplings exist even on toroidal backgrounds; and second to
demonstrate that the resulting $E2$-instanton effects do have the
potential to yield Majorana mass terms for the right-handed
neutrinos within the range $10^8-10^{15}$ GeV.

\subsection{ Background on the $T^6/\Z_2\times \Z_2'$ orientifold}
\label{O-background}
Consider the orientifold $T^6/\Z_2\times \Z_2'$
with Hodge numbers $(h_{11}, h_{12}) = (3,51)$. We stick to the
notation of \cite{Blumenhagen:2005tn}, to which we refer for details
of the geometry and the construction of rigid cycles. The orbifold
group is generated by $\theta$ and  $\theta'$ acting as reflection
in the first and last two tori, respectively. 

This background exhibits two types of factorizable special
Lagrangian three-cycles. The first class is given by the usual
non-rigid bulk cycles \bea \Pi_a^B = 4 \, \bigotimes_{I=1}^3
\,(n_a^I [a^I] + \widetilde m_a^I [b^I]), \eea defined in terms of
the fundamental one-cycles $[a^I], [b^I]$ of the $I$-th $T^2$ and
the corresponding wrapping numbers $n_a^I$ and  $\widetilde m_a^I=
m_a^I + \beta^I n_a^I$. Here $\beta^I=0, 1/2$ for rectangular and
tilted tori, respectively.

In addition there exist so-called $g$-twisted three-cycles \bea
\Pi^g_{ij} = n^{I_g} [\alpha^g_{ij,n}] + \widetilde m^{I_g}
[\alpha^g_{ij,m}], \eea where ${i,j} \in \{1,2,3,4\} \times
\{1,2,3,4\}$ labels one of the 16 blown-up fixed points of the
orbifold element $g  = \theta, \theta', \theta \theta' \in
\Z_2\times \Z_2'$. The cycles $[\alpha^g_{ij,n}]$ (
$[\alpha^g_{ij,m}]$) can be understood as twice the product of the
corresponding ${\mathbb P}_1$ and the one-cycle $[a]^{I_g}$
($[b]^{I_g}$) in the $I_g$-th $T^2$ invariant under $g$. Here $I_g =
3,1,2$ for $g  = \theta, \theta', \theta \theta'$, respectively.

These twisted cycles are the building blocks for certain fractional
cycles $\Pi^F$ charged under all three twisted sectors. They are
rigid and will serve as candidates for $E2$-branes contributing to
the superpotential. The general expression for $\Pi^F$ is given by
\bea \label{fraccyc} \Pi^F = \frac{1}{4} \Pi^B + \frac{1}{4} \Bigl(
\sum_{i,j \in S_{\theta}} \epsilon^{\theta}_{ij} \Pi^{\theta}_{ij}
\Bigr)+ \frac{1}{4} \Bigl(  \sum_{j,k \in S_{\theta'}}
\epsilon^{\theta'}_{jk} \Pi^{\theta'}_{jk}  \Bigr) + \frac{1}{4}
\Bigl(  \sum_{i,k \in S_{\theta \theta'}} \epsilon^{\theta
\theta'}_{ik} \Pi^{\theta \theta'}_{ik}  \Bigr). \eea The sets $S_g$
denote the four different fixed points in the $g$-twisted sector
compatible with the bulk wrapping numbers and the concrete position
of the brane, as detailed in \cite{Blumenhagen:2005tn}. A given set
of bulk wrapping numbers allows for a choice of $2 \times 2 \times 2
=8$ inequivalent positions of the fractional brane. Each of these
branes is further specified by the signs $\epsilon^g_{ij}$,
corresponding to the orientation with which the various ${\mathbb
P}_1$ are wrapped in the twisted sector. They are subject to various
consistency conditions \cite{Blumenhagen:2005tn} such that for each
choice of position of the fractional brane, there are only $8$
inequivalent choices of  $\epsilon^g_{ij}$.

\begin{figure}
\begin{center}
\includegraphics[scale=0.65]{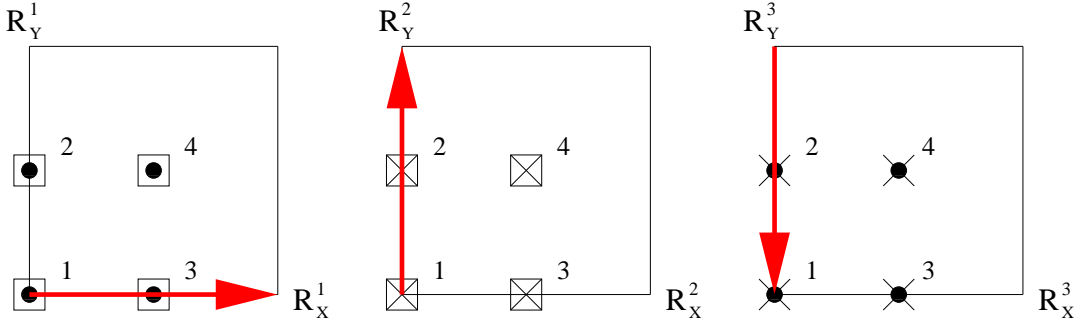}
\caption{$T^6/\Z_2\times \Z_2'$ with $\beta^1=\beta^2= \beta^3=0$.} \label{fig1}
\end{center}
\end{figure}

The orientifold action
$\Omega\mathcal{R}$ on the untwisted cycles follows from
\begin{align}
\Omega\mathcal{R}: \,[a_I]\rightarrow [a_I] \qquad
\Omega\mathcal{R}:\, [b_I]\rightarrow -[b_I],
\end{align}
whereas the twisted cycles transform as
\begin{align}
\label{Omegatwisted}
\Omega\mathcal{R}: \,\alpha^g_{ij,n}\rightarrow-
\eta_{\Omega\mathcal{R}}\,\eta_{\Omega\mathcal{R}g} \,
\alpha^g_{\mathcal{R}(i)\mathcal{R}(j),n},\qquad \Omega\mathcal{R}:
\,\alpha^g_{ij,m}\rightarrow
\eta_{\Omega\mathcal{R}}\,\eta_{\Omega\mathcal{R}g} \,
\alpha^g_{\mathcal{R}(i)\mathcal{R}(j),m}.
\end{align}
Here the reflection $\mathcal{R}$ leaves all
fixed points of an untilted two-torus invariant and acts on the fixed
points in a tilted two-torus as
\begin{align}
\mathcal{R}(1)=1 \qquad \mathcal{R}(2)=2 \qquad\mathcal{R}(3)=4
\qquad\mathcal{R}(4)=3.
\end{align}
The signs $\eta_{\Omega\mathcal{R}g}= \pm 1$ defining the
orientifold action are subject to the constraint
\begin{align}
\label{etaconst}
\eta_{\Omega\mathcal{R}}\,\eta_{\Omega\mathcal{R}\theta}\,
\eta_{\Omega\mathcal{R}\theta'}\,\eta_{\Omega\mathcal{R}\theta\theta'}= -1.
\end{align}
In our subsequent example we choose for simplicity all tori to be untilted and
\begin{align}
\label{eta-choice}
\eta_{\Omega\mathcal{R}}= \eta_{\Omega\mathcal{R}\theta}
=\eta_{\Omega\mathcal{R}\theta\theta'}=
-\eta_{\Omega\mathcal{R}\theta'}=1.
\end{align}
In this case, the orientifold image of the cycle $\Pi^F$ in equ.
(\ref{fraccyc}) is given by $\Pi'^F$, \bea \label{fraccyc'} \Pi'^F =
\frac{1}{4} \widehat\Pi^B -  \frac{1}{4}\,  \Bigl(  \sum_{i,j \in
S_{\theta}} \epsilon^{\theta}_{ij} \widehat\Pi^{\theta}_{ij}  \Bigr)
+ \frac{1}{4} \, \Bigl(  \sum_{j,k \in S_{\theta'}}
\epsilon^{\theta'}_{jk}
\widehat\Pi^{\theta'}_{jk}  \Bigr) - \frac{1}{4}
\Bigl(  \sum_{i,k \in S_{\theta \theta'}} \epsilon^{\theta
\theta'}_{ik} \widehat\Pi^{\theta \theta'}_{ik}  \Bigr) \nonumber, \eea where the $\, \, \widehat{}\, \, $
denotes the substitution $m^I\to -m^I$ (see also
\cite{Blumenhagen:2006ab}).

\noindent The fixed point locus sets expressed in terms of the toroidal cycles
take the form
\begin{align}
\pi_{O6}=2\,
[a_1]\,[a_2]\,[a_3]- 2 \,
[b_1]\,[b_2]\,[a_3]+ 2\,
[a_1]\,[b_2]\,[b_3]- 2\,
[b_1]\,[a_2]\,[b_3].
\end{align}

We also recall the topological intersection number
$I_{ab}$ of two  bulk branes $\Pi_a^B$
and $\Pi_b^B$,
\begin{align}
\label{Iab}
I_{ab}=4\prod^{3}_{i=1} \,(n^i_a\,m^i_b-n^i_b\,m^i_a).
\end{align}
Since our conventions are such that a stack of $N_a$ coincident
branes away from the orientifold carries gauge group $U(N_a/2)$ upon
taking the ${\mathbb Z}_2$-projection on the Chan-Paton factors into
account,  the quantity (\ref{Iab}) counts the number of chiral
multiplets in the bifundamental of the gauge group $({\bf \overline
{\frac{N_a}{2}}}, {\bf  \frac{N_b}{2}})$ living at the intersection
of two stacks of $N_a$ and $N_b$ bulk cycles $a$ and $b$,
respectively.
The number of chiral multiplets transforming as antisymmetric and
symmetric representations under $U(N_a/2)$ is
\begin{align}
I^{Anti}=\frac{1}{2}(I_{a'a}+I_{O6a}), \qquad \qquad
I^{Sym}=\frac{1}{2}(I_{a'a}-I_{O6a}),
\end{align}
where
\bea
I_{aa'}&=&-32\,n^1_a\,m^1_a\,n^2_a\,m^2_a\,n^3_a\,m^3_a,  \nonumber\\
I_{O6a}&=& 8\,\,m^1_a\,m^2_a\,m^3_a
-8\,\,n^1_a\,n^2_a\,m^3_a +
8\,\,m^1_a\,n^2_a\,n^3_a
-8\,\,n^1_a\,m^2_a\,n^3_a\,\, \nonumber.
\eea
In our applications the $E2$-instanton will wrap a rigid cycle $\Xi$. Its
intersection with a bulk brane $\Pi_a^B$ and its image $(\Pi^B_a)'$ is independent of the twisted charge of $\Xi$,
\begin{align}
I_{\Xi a}=\prod^3_{i=1} (n^i_{\Xi}\, m^i_a-n^i_{a}\,m^{i}_{\Xi}), \qquad
I_{\Xi a'}=-\prod^3_{i=1} (n^i_{\Xi}\, m^i_a+n^i_{a}\,m^{i}_{\Xi}).
\end{align}

\subsection{Wrapping numbers and spectrum of a local brane set-up}

We proceed with the construction of a local $SU(5)$ GUT-like
model. In this approach\footnote{See e.g.
\cite{Cvetic:2002pj,Chen:2006sd} for global constructions of a
similar type. }, the Standard Model arises from a stack of  $10$
coincident $D6$-branes carrying gauge group $U(5)= SU(5) \times
U(1)$, where the abelian part is massive due to the Green-Schwarz
mechanism. For simplicity, we choose the GUT stack to be given by
non-rigid bulk branes so that the GUT group can be broken down to
the Standard Model gauge group by invoking brane-splitting, i.e. by
giving suitable VEVs to the GUT Higgs fields in the adjoint of
$SU(5)$. Right-handed neutrinos are localized at the intersection of
two more stacks $a$ and $b$ of D-branes such that they are indeed
singlets under the GUT $SU(5)$. We choose  $a$ and $b$ to be
likewise given by bulk cycles. The actual "Standard Model" spectrum
arises, upon GUT breaking, from 4 chiral generations in the ${\bf
10}$ of $SU(5)$ as well as chiral multiplets transforming as ${\bf
\overline 5}$ localized at the intersection of $c$ and $a$. The
electroweak Higgs field candidates ${\bf 5}_H$
arise from the  intersection between stack $c$ and $b$.
In table \ref{wrapping number} we display the wrapping
numbers of the  stacks $a,b$ and $c$ of D-branes in a
particular realization of the described local set-up. This table
also contains the intersection numbers of the stacks and their image
stacks with the $E2$-instanton to be defined later in equ.
(\ref{E2}). Table \ref{spectrum} gives the
multiplicities of the "Standard Model" spectrum. In addition, there is chiral exotic matter which we do not make explicit. 
\begin{table}[htb]
\begin{center}
\begin{tabular}{|c||c|c|c|}
    \hline \rm{stack} & $N$ & $(n^1,m^1)\times (n^2,m^2)\times
(n^3,\tilde{m}^3)$ & $I_{E2x}$ \\
\hline \hline $E_2$ & 1 & $(1,0)\times (0,1)\times (0,-1)$ & \\
\hline \hline
    $a$ &  2 & $(1,2)\times (1,1)\times (-1,-1)$  &2 \\  \hline
    $b$&  2  & $(-3,-2)\times (1,-1)\times (-1,0)$&-2 \\  \hline
    $c$&  10  & $(1,0)\times (-1,2)\times (-2,-3)$&0 \\
\hline
\end{tabular}
\end{center}
\caption{\label{wrapping number} Wrapping numbers of the local
setup.}
\end{table}

Note that the charges $(-1_a,1_b)$ of $N_R^c$  under the global
symmetries $U(1)_a$ and $U(1)_b$ indeed  forbid perturbative
Majorana masses. We therefore seek to generate such terms
non-perturbatively. In order for the potential instanton-induced
Majorana mass terms to yield, via the standard see-saw mechanism,
hierarchically small masses for the neutrino mass eigenstates, the
model has to allow for perturbatively generated Dirac neutrino
masses.
\begin{table}[h!]
\begin{center}
\begin{tabular}{|c|c|c|c|}
    \hline \rm{sector} & $I_{xy}$ & representation&     matter\\
\hline \hline
  $(c,c')$ & $4$
  &
  Antisym
  &
  $\textbf{10}$\\
  \hline
    $(c,a)$ & $24$ & $(\overline{c},a)$  & $\overline {\textbf{5}}$\\  \hline
     $(c,b)$& $-24$ & $(c, \overline{b})$ &   $\textbf{5}_H$ \\  \hline
    $(a,b)$& $32$ & $(\overline{a},b)$  &  $N^c_R$  \\  \hline
\end{tabular}
\end{center}
\caption{\label{spectrum}Matter spectrum of the local setup.}
\end{table}

\noindent This feature is indeed realized, as can be seen from the
concrete intersection pattern in table \ref{spectrum}.
The Dirac mass terms are encoded schematically in the coupling \bea
\label{pertDirac} H \, L_L\, N_R^c \in  {\bf 5}_H \,\, {\bf \overline
5}\, \, \, {\bf 1}. \eea 
The magnitude of Dirac mass terms depends on the magnitude of the above
Yukawa couplings and the vacuum expectation value of the Standard Model
Higgs fields $H$.  Yukawa couplings are determined by a tree-level disk
amplitude calculation \cite{Cremades:2003qj,Cvetic:2003ch}. They are in general exponentially suppressed
by the area of the leading triangle formed by the intersecting branes and
thus depend on the K\"ahler and open string moduli of the background in
the way found in \cite{Cremades:2003qj,Cvetic:2003ch}.  For a toroidal-type set-up, constraints on the
four-dimensional gauge couplings and Planck length  $\ell_{Planck}$
typically constrain $\ell_s\sim \ell_{Planck}$ and thus  
the terms in the leading exponents typically  cannot be larger than
$\cal{O}$(10).  For an early concrete analysis of
these suppression terms for the first globally consistent three-family
supersymmetric Standard-like model \cite{CSUI,CSUII}, see \cite{CLS}.  
The analysis of Dirac mass terms can  be repeated in a straightforward
way  for our  set-up  by parallel splitting of GUT $SU(5)$ branes into
three SM  branes and analysing  the associated  Yukawa couplings for
quarks and leptons.   In general Dirac neutrino masses
will follow  the mass pattern of charged leptons and quarks, and are thus
in the  same mass range.

In order for the model to be supersymmetric each stack  of branes
has to satisfy the two conditions  \cite{Blumenhagen:2005tn}
\begin{align}
m^1_x\,m^2_x\,m^3_x-\sum_{I\neq J\neq K}\frac{n^I_x\,n^J_x\,
m^K_x}{U^{I}\,U^{J}}=0
\end{align}
and
\begin{align}
n^1_x\,n^2_x\,n^3_x-\sum_{I\neq J\neq K}m^I_x\,m^J_x\,
n^K_x\,U^{I}\,U^{J}>0\,\,,
\end{align}
where $U^I$ denotes the complex structure modulus $U^I=R^I_Y/R^I_X$
of the $I-th$ torus with radii $R^I_X,R^I_Y$. The brane set-up
satisfies the equations above for the following choice of complex
structure moduli $U^I$,
\begin{align}
U^{1}= \sqrt{3}\,\,, \qquad \qquad
U^{2}=\frac{2}{\sqrt{3}}\,\,, \qquad \qquad
U^{3}=\frac{8}{3 \sqrt{3}}\,\,. \label{complex structure moduli}
\end{align}
We stress once more that the brane configuration of table
\ref{wrapping number} as such does not satisfy all of the tadpole
cancellation conditions ensuring global consistency and therefore
only represents a local model. In particular, its spectrum is not anomaly-free.

\subsection{The $E2$-instanton}
\label{inst}

We are now in a position to analyze the $E2$-instanton sector of the
local model defined in the previous section. We are particularly
interested in fermion bilinears of type (\ref{bilinear_final}) for
the right-handed neutrinos. As described in detail, they are due to
$E2$-instantons wrapping a rigid supersymmetric cycle $\Xi$ subject to (\ref{intnum_e2}) such that the $\overline \theta_{\dot \alpha}$ modes are projected out \cite{LA} and which do
not give rise to zero modes  in the $\Xi-\Xi'$ sector.  
Our analysis
therefore consists in two steps: First classify the rigid cycles
$\Xi$ with no $\Xi-\Xi'$-modes and then distinguish them according
to their charged zero modes structure. As it will turn out, in our setup the first step automatically guarantees absence of the $\overline \theta_{\dot \alpha}$ modes as well.

The only type of sLags under technical control corresponds to the
class of factorizable three-cycles described in section
\ref{O-background}. Even though a complete analysis of the instanton
sector should take into account all possible sLags our analysis is
therefore forced to content itself with this special class. A closer
look reveals that the constraint $\Xi \cap \Xi' =0$ is extremely
restrictive and can be met (at best) by two different types of rigid
cycles. Either $\Xi$ and its orientifold image $\Xi'$ are parallel,
but separated in at least one of the three tori. In that case the
vector-like fermionic zero modes in the  $\Xi-\Xi'$ sector carry a
mass proportional to the distance between $\Xi$ and $\Xi'$.
Alternatively, one can consider those rigid cycles with the property
$\Xi=\Xi'$. For such invariant cycles, the massless modes in the
$E2-E2'$ sector are identical to the geometric moduli of the cycle.
Rigidity then guarantees the absence of open string moduli and
$E2-E2'$ zero modes at the same time.

As is immediately clear, the explicit form of the orientifold action
on the fixed points in the ${\mathbb Z}_2 \times {\mathbb Z}'_2$
background at hand excludes the first type of cycles. Potential
candidate cycles for the second class have to lie on top of one of
the four orientifold planes to ensure that the bulk part is indeed
mapped to itself under $\Omega {\mathcal R}$. In addition, we have
to take into account the non-trivial orientifold action on the
$g$-twisted sector encoded in (\ref{Omegatwisted}). Depending on the
choice of $\eta_{\Omega {\cal R} g}$, a certain combination of
twisted charges of $\Xi$ may also be $\Omega$ invariant such that
$\Xi=\Xi'$. With our given choice (\ref{eta-choice}) for
$\eta_{\Omega {\cal R} g}$, only those rigid cycles  parallel to the
$x^1$-, $y^2$- and $y^3$-axis have a chance to be invariant. This
can be seen e.g. from the fact that the $\alpha^g_{ij,m}$ are
invariant only for $g=\theta$ and $g= \theta \theta'$, cf. equ.
(\ref{Omegatwisted}).
Due to the additional minus sign in the orientifold projection
resulting from the external DD boundary conditions, the
non-dynamical gauge group on a stack of $N$ such invariant
instantons is $SO(N)$\footnote{Recall that for $D6$-branes wrapping
invariant cycles, the gauge group was determined in
\cite{Blumenhagen:2005tn} to be $Sp(2N)$.}. In particular, this means that we can wrap a \emph{single} instanton on this invariant cycle and that the orientifold projection removes the unwanted $\overline \theta_{\dot \alpha}$ modes from the $E2-E2$ sector.

To completely specify a cycle of this type we have
to choose the explicit values for the bulk wrapping numbers, the
actual position of the brane and thus the fixed points $S_g$ to be
wrapped in the twisted sector and finally the $\epsilon^g$ signs.

We start with the bulk wrapping numbers.

From (\ref{intnum_e2}) we need intersection numbers  $I_{E2 a}=2$
and  $I_{E2 b}=-2$ in order for the instanton to exhibit abelian
charges $Q_a=2$ and  $Q_b=-2$. Since $E2=E2'$, the zero modes in the
$E2-a$ and $a'-E2$ are identified and do therefore not count as
independent. This uniquely determines the bulk wrapping numbers of
the fractional cycle to
\begin{align}
\label{bulkwrapping}
\Pi^B_{\Xi}\,:\,\,\, [(1,0)\,(0,1)\,(0,-1)].
\end{align}

We note in passing that by this analysis there exist no
$E2$-instantons leading to dangerous open string tadpoles of the
form $\Phi \, e^{-S_{inst}}$ for matter between the stacks $a$, $b$
or $c$. For perturbatively well-defined string vacua such tadpoles
would spoil stability  at the quantum level.

Given the bulk wrapping numbers (\ref{bulkwrapping}), we have the
following options for the twisted sector: The fractional brane can
run through the fixed points $(1,3)$ or $(2,4)$ in the first torus
and through $(1,2)$ or $(3,4)$ in the second and third torus (see figure \ref{fig1}).

Thus, we have $8$ different positions for the invariant
cycle, together with the mentioned $8$ inequivalent sign choices
$\epsilon^g$ for each position. One example of these $64$ different
cycles takes the form
\begin{align}
\label{E2}
\Pi_{\Xi}=\frac{1}{4}\Pi^B_{\Xi}-\frac{1}{4}\sum_{i,j\epsilon(13)\times(12)}
\alpha^{\theta}_{ij,m}+ \frac{1}{4}\sum_{j,k\epsilon(12)\times(12)}
\alpha^{\theta'}_{jk,n}+ \frac{1}{4}\sum_{i,k\epsilon(13)\times(12)}
\alpha^{\theta\theta'}_{ik,m}.
\end{align}
It corresponds to an $E2$ passing through the origin in each of the
three tori and the choice $\epsilon^g_{ij}=1$ in all sectors, as
depicted in figure \ref{fig1}. The remaining 63 instanton cycles are
obvious modifications of this one. One may convince oneself that the
choice (\ref{eta-choice}) indeed yields $\Xi=\Xi'$, thus qualifying
$\Xi$ as an $E2$-instanton cycle relevant for the superpotential.

\noindent In the sequel, when analyzing the single $E2$-instanton
sector relevant for the Majorana mass terms, we have to consider
each of these inequivalent choices of the twisted sector. The final
result for the non-perturbative coupling will be the sum of the
contribution from each sector. Our explicit computation will be for
instanton (\ref{E2}) and we will discuss the remaining contributions
at the end of section \ref{computation_maj}.

It is crucial for the generation of fermion bilinears that there
exist no \emph{chiral} zero modes from strings stretching between
the $E2$-instanton and the stack $c$ since $I_{\Xi c}=0$. However,
since $\Xi$ and $c$ share the same bulk wrapping numbers in the
first torus, there exist vector-like pairs at the intersection of
$\Xi$ and $c$ in the second and third torus with mass proportional
to twice the distance between $\Xi$ and $c$ in the first torus. In
order to avoid \emph{massless} vector-like pairs, we have to assume
that the latter stack is separated from the instanton in the second
torus by a non-zero distance. In the absence of effects stabilizing
the open string moduli, we can freely move along the  corresponding
flat direction in moduli space. \\ To summarize, the zero mode
structure meets the required constraints to give rise to Majorana
mass terms for the right-handed neutrinos $\nu_R^c$ sitting in the
superfields $N_R^c$ at the intersection of branes $(a,b)$ carrying
abelian charge $(-1_a, 1_b)$ (see table \ref{spectrum}). One may
check that $U(1)_a$ and $U(1)_b$ are indeed both broken as a gauge
symmetry since the corresponding vector potentials acquire a
St\"uckelberg-type mass. Recall that this is the \emph{conditio sine
qua non} for the instanton to off-set the abelian charge violation
of the open string operator in the non-perturbative coupling.

\subsection{Computation of the Majorana masses }
\label{computation_maj} We finally apply the results of section
\ref{sec_Ampl-CFT} and obtain the  neutrino Majorana mass terms by
evaluating the two-point correlator \bea
\label{majoranamassamplitude} <\nu^A \nu^B>_{E2} &=&
-\frac{1}{2!}\,\frac{{\cal V}_{E2}}{16} \frac{g_s}{2\pi}\,\int d^4  x_E \,\int
d^2  \theta \, \int d^2 \lambda_a\, \int d^2 \overline\lambda_{b}\,
\, e^{-S_{inst.}}\, e^{Z'} \nonumber
\\ && \sum_{i,j,k,l} < V^{-\frac{1}{2}}_{\Theta^{\alpha}}  V^{-\frac{1}{2}}_{\overline\lambda^k_{b}} V^{-\frac{1}{2}}_{{\nu}^A_{\alpha}} V^{-\frac{1}{2}}_{
\lambda^i_a}
>\,< V^{-\frac{1}{2}}_{\Theta^{\beta}}  V^{-\frac{1}{2}}_{\overline\lambda^l_{b}} V^{-\frac{1}{2}}_{{\nu}^B_{\beta}} V^{-\frac{1}{2}}_{
\lambda^j_a}
>. \label{majorana masses}\eea
For the concrete intersection anglese
\begin{align}
\theta^{1}_{ab}&=0.86,\,\,\,\,\,\,\,\, \qquad \theta^{2}_{ab}=-0.54,
\,\,\,\,\,\,\,\,\,\,\,\,\,\,\,
\theta^{3}_{ab}=-0.32, \nonumber\\
\theta^{1}_{E2a}&=0.41, \, \,\,\,\,\,\,\,\,\,\,\,\,\,\,\,
\theta^{2}_{E2a}=-0.23,\quad\,\,\,\,\:
\theta^{3}_{E2a}=-0.18,\\
\nonumber \theta^{1}_{E2b}&=-0.73, \qquad \theta^{2}_{E2b}=-0.77,
\qquad \theta^{3}_{E2b}=-0.50
\end{align}
the vertex operators read \begin{align} \nonumber
 V_{\nu}
 &=\ell_s^{\frac{3}{2}}\, \Omega_{ba}\, {\nu}_{\alpha}
\, S^{\alpha}(z) \, \sigma_{1-\theta^1_{ab}}(z)\,
e^{-i(\theta^1_{ab}-\frac{1}{2}) H_1(z)} \qquad \qquad\qquad
\qquad\qquad
\\& \qquad \qquad \qquad \qquad \times \prod_{I=2}^3 \sigma_{-\theta^I_{ab}}(z)\,
e^{-i(\theta^I_{ab}+\frac{1}{2}) H_I(z)}\, e^{i k_{\mu}X^{\mu}(z)}
\,e^{-\varphi(z)/2}, \, \nonumber\\
 V_{\lambda_{a}} &=
\Omega_{aE2} \,\lambda_a \, \Sigma(z)\, \,
\sigma_{1-\theta^1_{E2a}}(z)\, e^{-i(\theta^1_{E2a}-\frac{1}{2})
H_1(z)}
 \qquad \qquad\qquad \qquad\qquad
\\ & \qquad \qquad\qquad \qquad \times \prod_{I=2}^3 \sigma_{-\theta^I_{E2a}}(z)\,
e^{-i(\theta^I_{E2a}+\frac{1}{2}) H_I(z)}\,
e^{-\varphi(z)/2},\nonumber\\
&\,\,\,\,\: V_{\overline\lambda_{b}} = \Omega_{E2b} \,
\overline\lambda_b
 \, \Sigma(z)\,\prod^{3}_{I=1}  \sigma_{1+\theta^I_{E2b}}(z)\,
e^{i(\theta^I_{E2b}+\frac{1}{2}) H_I(z)}\, e^{-\varphi(z)/2}
\nonumber.\end{align} It follows that the angle dependence of the
disk amplitude \bea < V^{-\frac{1}{2}}_{\Theta^{\alpha}}
V^{-\frac{1}{2}}_{\overline\lambda^k_{b}}
V^{-\frac{1}{2}}_{\nu^A_{\alpha}} V^{-\frac{1}{2}}_{
\lambda^i_a}>\,=\frac{2\pi}{g_s}\,\ell_s^{\frac{3}{2}}\, C^A_{ik}\,
\,e^{-ik^A_{\mu}x^{\mu}_0} \quad (\theta^{\alpha} \overline
\lambda^k_b    {\nu}^A_{\alpha} \,\lambda^i_a )\,\, \label{}\eea is
given by
 \bea
C^A_{ik}=\pi\left[4\pi\,\Gamma_{1-\theta^1_{ab},1-\theta^1_{E2a},1+\theta^1_{E2b}}\,\prod^3_{I=2}
4\pi\,\Gamma_{-\theta^I_{ab},-\theta^I_{E2a},1+\theta^I_{E2b}}\right]^{\frac{1}{4}}
\sum_{m_j}e^{-{\cal A}^{A}_{ik}(m_j)} \,\, \label{constants}\eea for
index combinations with non-vanishing diagrams. Before turning to
this question, we first investigate the instanton suppression factor
\bea \label{tree-supp} e^{-S_{inst}} = e^{-{2\pi\over \ell_s^3\,
g_s} {\rm Vol}_{E2} } = e^{-{2\pi\over \alpha_{\rm GUT}} { {\rm
Vol}_{E2}\over
              {\rm Vol}_{\Pi_c} }},
\eea where the last equation uses the standard relation (see
e.g.\cite{Klebanov:2003my}) \bea \alpha_{GUT} = g_s \frac{{\rm
Vol}_{\Pi_c}}{\ell_s^3} \eea for ${\alpha}_{GUT}$ in terms of the
volume of the GUT stack $c$. Given the geometric data of our
concrete string vacuum, the ratio ${\rm Vol}_{E2}\over {\rm
Vol}_{\Pi_a}$ can easily be computed and is determined entirely by
the wrapping numbers in table \ref{wrapping number} and the complex
structure moduli \eqref{complex structure moduli}, \bea \frac{{\rm
Vol}_{E2}}{{\rm Vol}_{\Pi_c}} = \Bigl(\prod_I \frac{(n^I_{E2})^2 +
(\widetilde m^I_{E2})^2 U_I^2}{(n^I_{c})^2 + (\widetilde m^I_{c})^2
U_I^2}\Bigr) ^{1/2} =\frac{8}{57}\,\,. \eea

\begin{figure}
\begin{center}
\includegraphics[scale=0.65]{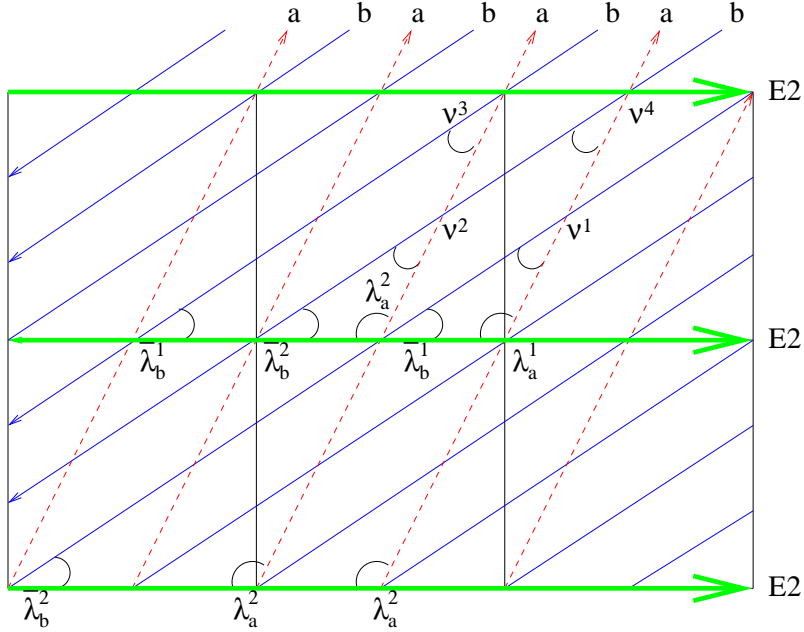}
\caption{Intersection pattern in first torus.}\label{diracneutrino}
\end{center}
\end{figure}

As discussed in section \ref{sec_Fam}, for  $C^A_{ik}$ to be
non-vanishing in each torus the modes $\lambda^{i}_a$,
$\overline\lambda^k_{b}$ and ${\nu}^A$ have to form a triangle. Let
us analyze this nontrivial constraint for our setup. Figure
\ref{diracneutrino} displays the intersection in the first torus.

One can easily read off that the
combinations of $\lambda^{i}_a$, $\overline\lambda^k_{b}$ and
${\nu}^A$ with triangles in the first torus have the
same structure as in the example in section \ref{sec_Fam}.\\
\begin{figure}
\begin{center}
\includegraphics[scale=0.65]{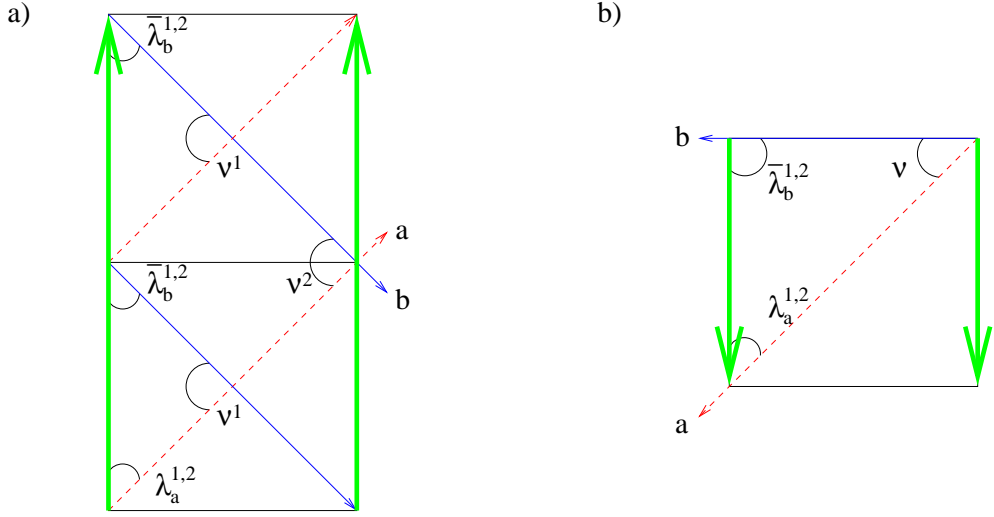}
\caption{Intersection pattern in second and third torus.}\label{resttori}
\end{center}
\end{figure}
The remaining two tori are depicted in figure \ref{resttori}.
While $a$ and $b$ intersect twice in the second torus there is only
one intersection in the third one. Most importantly, the replication
of $\lambda_a$ and $\overline \lambda_b$ modes is entirely due to
multiple intersections in the first torus.

The complete location of a neutrino ${\nu}^{i,j}$ is described by
two upper indices $i$ and $j$, where $i$ denotes the position in the
first torus while $j$ gives the location in the
second\footnote{Recall that $a$ and $b$ are bulk branes and so that
each intersection point gives rise to 4 right-handed neutrinos
$\nu_R^c \equiv \nu$. This yields the overall $4 \times 8 =32$ of
them. We leave the additional factor of $4$ implicit. }. Ignoring
all higher worldsheet instanton effects we obtain
\bea
\label{finalMajoranamess}
 <\nu^A \nu^B>_{E2} &=& \frac{2 \pi \,{\cal V}_{E2}}{g_s}\,\,
\overrightarrow{v}^T\,{\cal M}\,\overrightarrow{v} \,\,(2\pi)^4
\delta^4(k^A+k^B)\,\,.
 \eea
 Here $\overrightarrow{v}$ is defined as \bea
\overrightarrow{v}^T=({\nu}^{1,1},\,{\nu}^{2,1},\,
{\nu}^{3,1},\,{\nu}^{4,1},\,{\nu}^{1,2},\,
{\nu}^{2,2},\,{\nu}^{3,2},\, {\nu}^{4,2}) \eea and the
 $8 \times 8$ matrix ${\cal M}$  takes the form
\begin{align}
{\cal M} =x\,M_s\, e^{-{16\pi \over 57 \alpha_{\rm GUT}}}\,
\left(
\begin{array}{cccc}
A & 0 & B & 0 \\
0 & C & 0& D\\
B & 0 & E & 0\\
0&  D & 0&F
\end{array} \right)\,\,,
\label{majorana mass matrix}
\end{align}
where $x$ is given by \bea
x=\frac{\pi^2}{16}\left[4\pi\,\Gamma_{1-\theta^1_{ab},1-\theta^1_{E2a},1+\theta^1_{E2b}}\,\prod^3_{I=2}
4\pi\,\Gamma_{-\theta^I_{ab},-\theta^I_{E2a},1+\theta^I_{E2b}}\right]^{\frac{1}{2}}\,\,e^{Z'}\,\,.
\eea At 'tree-level', i.e. ignoring the corrections  due to the one-loop determinant $e^{Z'}$, the numerical
factor is approximately $x \approx 0.87$. The building blocks of
${\cal M}$ in \eqref{majorana mass matrix} take a similar form as in
the simpler example in section \ref{sec_Fam},
\begin{align}
A&=\left(
\begin{array}{cc}
e^{-(\alpha+\beta+2\kappa+2\tau)}&\frac{1}{2}\left(e^{-2(\alpha+\kappa+\tau)}+e^{-2(\beta+\kappa+\tau)}\right)\\
\frac{1}{2}\left(e^{-2(\alpha+\kappa+\tau)}+e^{-2(\beta+\kappa+\tau)}\right)
&e^{-(\alpha+\beta+2\kappa+2\tau)}
\end{array}\right)\,\,\\
B&=\left(
\begin{array}{cc}
e^{-(\alpha+\beta+\kappa+\mu+2\tau)}&\frac{1}{2}\left(e^{-(2\alpha+\kappa+\mu+2\tau)}+e^{-(2\beta+\kappa+\mu+2\tau)}\right)\\
\frac{1}{2}\left(e^{-(2\alpha+\kappa+\mu+2\tau)}+e^{-(2\beta+\kappa+\mu+2\tau)}\right)
&e^{-(\alpha+\beta+\kappa+\mu+2\tau)}
\end{array}\right)\,\,,
\label{matrices}
\end{align}
where $\alpha$, $\beta, \gamma$ and $\delta$ are defined in
\eqref{areas}, $\kappa$ ($\mu$) denotes the area of the triangle
spanned by $\nu^{i,1}$ $(\nu^{i,2})$, $\lambda_a$ and
${\overline\lambda}_b$ in the second torus, whereas $\tau$ is the
area in the third torus. The other $4$ building blocks can be easily
obtained in the following manner. Replacing $\alpha$ and $\beta$ in
$A$ ($B$) by $\gamma$ and $\delta$ yields $C$ ($D$), replacing in
addition  also $\kappa$ by $\mu$ one obtains $F$. In order to get
$E$ we just substitute in $A$ $\kappa$
by $\mu$.\\
The suppression due to worldsheet instantons depends crucially on
the open string moduli. However, since for a toroidal-type set-up the
four-dimensional Planck length $\ell_{Planck}$ is constrained
to be of the same order of magnitude as $\ell_s$,  the arguments of
the leading exponents typically range between zero and order  
one, and  thus these suppression factors are not excessive. In this case, also the factor
${\cal V}_{E2}$ in (\ref{finalMajoranamess}) is of order $1$.
In addition,  
for particular choices of open string moduli the area of triangles
vanishes and there is no suppression at all, e.g. for the instanton passing through the origin in each torus
and the choice of moduli as in figure (\ref{diracneutrino}, \ref{resttori}) there is
no suppression for the coupling $\nu^{3,2} \, \nu^{4,2}$. 

As discussed, the above coupling is the contribution of just $1$ out
of $8 \times 8$ rigid factorizable sLags with the required zero-mode
structure to yield Majorana mass terms. The first factor is due to
the two different positions of the $E2$-brane per two-torus,
corresponding to which of the fixed points it passes through (see
figure \ref{fig1}). Clearly, each of these $8$ choices comes with
different areas of the worldsheet triangles and therefore relative
suppression factors between families. For each geometric position we
have to sum in addition over $8$ inequivalent choices of signs of
the twisted charges $\epsilon^g_{ij}$. In our example with all other
$D6$-branes of the local model wrapping bulk cycles, the result for
the two-point coupling is independent of these twisted charges. In
particular, this is true for the one-loop determinant $e^{Z'}$
(\ref{1loop}) since the twist part of the $E2$ boundary state is
orthogonal to the boundary states of the bulk $D6$-branes and the
cross-cap. It follows that each of the $8$ geometrically distinct
sectors just contributes with an additional factor of $8$, i.e. the
various contributions from the factorizable rigid $E2$-instantons
with appropriate zero modes do add up to a non-vanishing result.
This is a fortunate result since in principle, one might have feared
non-trivial cancellations. Indeed, for heterotic $(0,2)$ models
explicit examples of such cancelations are known for special
constructions such as (half-)linear sigma models
\cite{Beasley:2003fx}, even though they do not correspond to the
generic situation. Of course, a complete classification of instanton
effects would require control over all special Lagrangian manifolds
and seems out of reach even on toridal backgrounds.

For a concrete choice of moduli, it is clear how to perform the sum over instanton configurations in detail.
In a crude approximation, this summation yields an additional factor of ${\cal O}(10)$.
With
$M_s=1.2\times10^{18}$ GeV and for $\alpha_{\rm GUT}$ within the
range of $\frac{1}{24}$ and $\frac{1}{20}$ \footnote{Note that the
familiar value $\alpha_{\rm GUT} \simeq \frac{1}{24}$ refers to the exact
MSSM spectrum. Given the large amount of exotic matter of our set-up we took a range of $\alpha_{GUT}$ values that is
somewhat larger that that of the MSSM. } the factor in \eqref{majorana mass matrix}
takes values in the range of $(0.1 - 1)\times 10^{11}$ GeV.
Therefore for the pattern of neutrino Dirac masses that are in
 the electroweak range (0.01-1) GeV, the  see-saw neutrino masses are in the
range  ($10^{-6}-0.1$) eV.

\section{Discussion}

In this paper, we have  continued the analysis of
\cite{Blumenhagen:2006xt} and provided the basic building blocks for
determining $E2$-instanton induced open string superpotential
couplings in toroidal Type IIA orientifolds. Specifically, we have
computed disk diagrams with insertion of one charged matter field
and appropriate instanton zero modes. These can then be combined
into non-perturbative $M$-point couplings. For the simplest case of
fermion bilinears, the exact result is given by
(\ref{bilinear_final}) and (\ref{C-const}). Such terms are of some
phenomenological interest since they can represent MSSM $\mu$-terms
or Majorana masses for right-handed neutrinos.

Focussing on the latter possibility, we have embedded $E2$-instanton
effects into a local toy model on the ${\mathbb Z}_2 \times {\mathbb
Z'}_2$ orientifold constructed such that the zero mode structure of
its instanton sector meets all requirements for the generation of
Majorana masses within the phenomenologically allowed window.
Together with perturbatively generated Dirac masses, these give
rise, via the see-saw mechanism, to hierarchically small neutrino
masses. The family mixing pattern among the various Majorana
couplings depends crucially on the relative suppression factor
governed, as for string tree-level Yukawa couplings,  by world-sheet
instanton effects \cite{Cremades:2003qj,Cvetic:2003ch}. A detailed
analysis of the resulting neutrino phenomenology in more realistic
models might be of some interest. Since our CFT results are directly
applicable also to non-perturbative MSSM $\mu$-terms, the
construction of at least local models featuring this effect might
also be worthwhile.

As we have described, within the tractable class of rigid
factorizable sLags on the ${\mathbb Z}_2 \times {\mathbb Z'}_2$
orientifold, the requirement of absence of zero modes in the
$E2-E2'$ sector singles out a small set of candidate cycles for the
instanton lying on top of one of the orientifold planes. This is a
major challenge for more realistic model building on toroidal
backgrounds. The main problem is that, to avoid unacceptable charged
zero modes between the instanton and other $D6$-branes beyond the ones
hosting the right-handed neutrinos, these branes have to be parallel
to the instanton and separated in one torus. As it turned out in the
cases considered, the use of such cycles makes it extremely hard to
satisfy all tadpole constraints in a supersymmetric set-up. All this
comes as no big surprise in view of the simple homology lattice of
toroidal backgrounds, and we do not expect these complications to be
insurpassable within the string landscape. In fact, finding M-theory
corners naturally incorporating such effects might serve as a
guideline in string model building.

\vskip 1cm
 {\noindent  {\Large \bf Acknowledgements}}
 \vskip 0.5cm
\noindent It is our great pleasure to thank  Ralph Blumenhagen for important discussions and comments on a draft of this article as well as Luis Ib{\' a}{\~ n}ez and Angel Uranga for correspondence on the Goldstone fermions. We also thank Vijay Balasubramanian, Volker Braun, Tamaz Brelidze, Michael Douglas, Paul Langacker, Luca  Mazzucato, Tao Liu, Diester L{\" u}st, Mike Schulz, Maximilian Schmidt-Sommerfeld, Jaemo Park and Erik Plauschinn for interesting conversations. 
This research was supported in part by the
Department of Energy Grant
DOE-EY-76-02-3071 and the Fay R. and Eugene L. Langberg
Endowed Chair.

\clearpage
\nocite{*}
\bibliography{rev}
\bibliographystyle{utphys}

\end{document}